\documentclass{aastex631}
\received{November 21st 2022}
\accepted{May 31st 2023}
\usepackage{chemformula}
\usepackage[version=4]{mhchem}

\newcommand{\sed}[1]{{#1}}

\shorttitle{Prebiosignatures}
\shortauthors{Claringbold et al.}

\begin{document}


\title{Prebiosignature Molecules Can Be Detected in Temperate Exoplanet Atmospheres with JWST}
\author{A. B. Claringbold}
\affiliation{Institute of Astronomy, University of Cambridge, Madingley Road, Cambridge CB3 0HA, UK}
\affiliation{Centre for Exoplanets and Habitability, University of Warwick, Gibbet Hill Road, Coventry CV4 7AL, UK}
\affiliation{Department of Physics, University of Warwick, Gibbet Hill Road, Coventry CV4 7AL, UK}
\author{P. B. Rimmer}
\affiliation{Cavendish Astrophysics, University of Cambridge, JJ Thomson Ave., Cambridge CB3 0HE, UK}
\affiliation{Department of Earth Sciences, University of Cambridge, Downing St., Cambridge CB2 3EQ, UK}
\author{S. Rugheimer}
\affiliation{Atmospheric, Oceanic and Planetary Physics Department, University of Oxford, Sherrington Road, Oxford OX1 3PU, UK}
\affiliation{Department of Physics and Astronomy, York University, 4700 Keele St., Toronto, ON M3J 1P3, Canada}
\author{O. Shorttle}
\affiliation{Institute of Astronomy, University of Cambridge, Madingley Road, Cambridge CB3 0HA, UK}
\affiliation{Department of Earth Sciences, University of Cambridge, Downing St., Cambridge CB2 3EQ, UK}


\begin{abstract}
    \noindent The search for biosignatures on exoplanets connects the fields of biology and biochemistry to astronomical observation, with the hope that we might detect evidence of active biological processes on worlds outside the solar system. Here we focus on a complementary aspect of exoplanet characterisation connecting astronomy to prebiotic chemistry: the search for molecules associated with the origin of life, prebiosignatures. Prebiosignature surveys in planetary atmospheres offer the potential to both constrain the ubiquity of life in the galaxy and provide important tests of current prebiotic syntheses outside of the laboratory setting. Here, we quantify the minimum abundance of identified prebiosignature molecules that would be required for detection by transmission spectroscopy using JWST. We consider prebiosignatures on five classes of terrestrial planet: an ocean planet, a volcanic planet, a post-impact planet, a super-Earth, and an early Earth analogue. Using a novel modelling and detection test pipeline, with simulated JWST noise, we find the detection thresholds of hydrogen cyanide (\ce{HCN}), hydrogen sulfide (\ce{H2S}), cyanoacetylene (\ce{HC3N}), ammonia (\ce{NH3}), methane (\ce{CH4}), acetylene (\ce{C2H2}), sulfur dioxide (\ce{SO2}), nitric oxide (\ce{NO}), formaldehyde (\ce{CH2O}), and carbon monoxide (\ce{CO}) in a variety of low mean molecular weight \sed{($<5$)} atmospheres. We test the dependence of these detection thresholds on \sed{M dwarf} target star and the number of observed transits, finding that a modest number of transits (1-10) are required to detect prebiosignatures in numerous candidate planets, including TRAPPIST-1e with a high mean molecular weight atmosphere. We find that the NIRSpec G395M/H instrument is best suited for detecting most prebiosignatures.

\end{abstract}

\section{Introduction}
With the recent launch of \textit{JWST} \citep[see][]{gardner2006james,rigby2022science}, we are on the cusp of characterising habitable exoplanets by probing their atmospheric structure, chemistry, and composition. Measuring the abundances of chemical species in exoplanet atmospheres could grant us an understanding of the geochemical and physical processes active on terrestrial exoplanets. Such species include molecules associated with life (biosignatures) and its origins (prebiosignatures). 
To explain the origin of life on Earth, a number of chemical syntheses have been explored that give rise to promising precursor molecules, such as the Miller-Urey experiment \citep{miller1953production} and the cyanosulfidic scenario \citep{patel2015common}, to name just two. Molecules that are prerequisites or products of these prebiotic chemistries, for instance hydrogen cyanide (\ch{HCN}), may be present in the atmospheres of exoplanets as trace species.
Detection of these prebiosignatures could provide insight into how life might arise both on our planet and in the universe \citep{rimmer2021life}, mitigating the sole example problem in our current understanding of the origin of life. The desire to find these prebiosignature molecules motivates us to explore the capability of \textit{JWST} to detect and constrain these signatures, and in particular to find the abundance threshold at which this should be possible. This can in turn inform both observing strategies, and the astronomical testability of prebiotic hypotheses, and hence connect the disparate fields of prebiotic chemistry and astronomy.

\subsection{Prebiosignatures}
Prebiosignatures are molecules that may be involved in prebiotic chemistry, as either the direct products or the feedstock of the pathways themselves (primary prebiosignatures) or molecules created by abiotic processes which may be involved in the origin of life: impacts, volcanism, stellar activity or lightning (secondary prebiosignatures). For example, the cyanosulfidic scenario uses hydrogen cyanide with UV radiation and a reducing sulfur species, either hydrogen sulfide \citep{patel2015common} or sulfur dioxide \citep{xu2018photochemical}, to generate RNA and protein precursors. The UV radiation requirements of this scenario place constraints on the astrophysical context in which it can occur.  \cite{ranjan2017surface} and \cite{rimmer2018origin} have explored these astrophysical limits, demonstrating how inactive, cool stars (M and late K dwarfs) have insufficient UV flux to drive this chemistry, although flares could mitigate this issue. The abundances of prebiotic species that arise in terrestrial exoplanet atmospheres could also place constraints on the astrophysical contexts of the origin of life, but are presently less well understood. Hence, there is a need to probe planetary atmospheres for traces of these molecules. 

The study of prebiosignatures is a natural extension of the search for biosignatures, and many works have explored the detection of life beyond Earth \citep[e.g.,][]{segura2005biosignatures,kaltenegger2010deciphering,rauer2011potential,krissansen2016detecting,seager2016toward,rugheimer2018spectra,konrad2022large,alei2022large,angerhausen2022large}. The nature of prebiosignatures and our (lack of) understanding of the origin of life draws us to take an open-minded approach, and explore any molecule or process which could support abiogenesis. 

\begin{figure}
    \centering
    
    \includegraphics[width=0.9\textwidth]{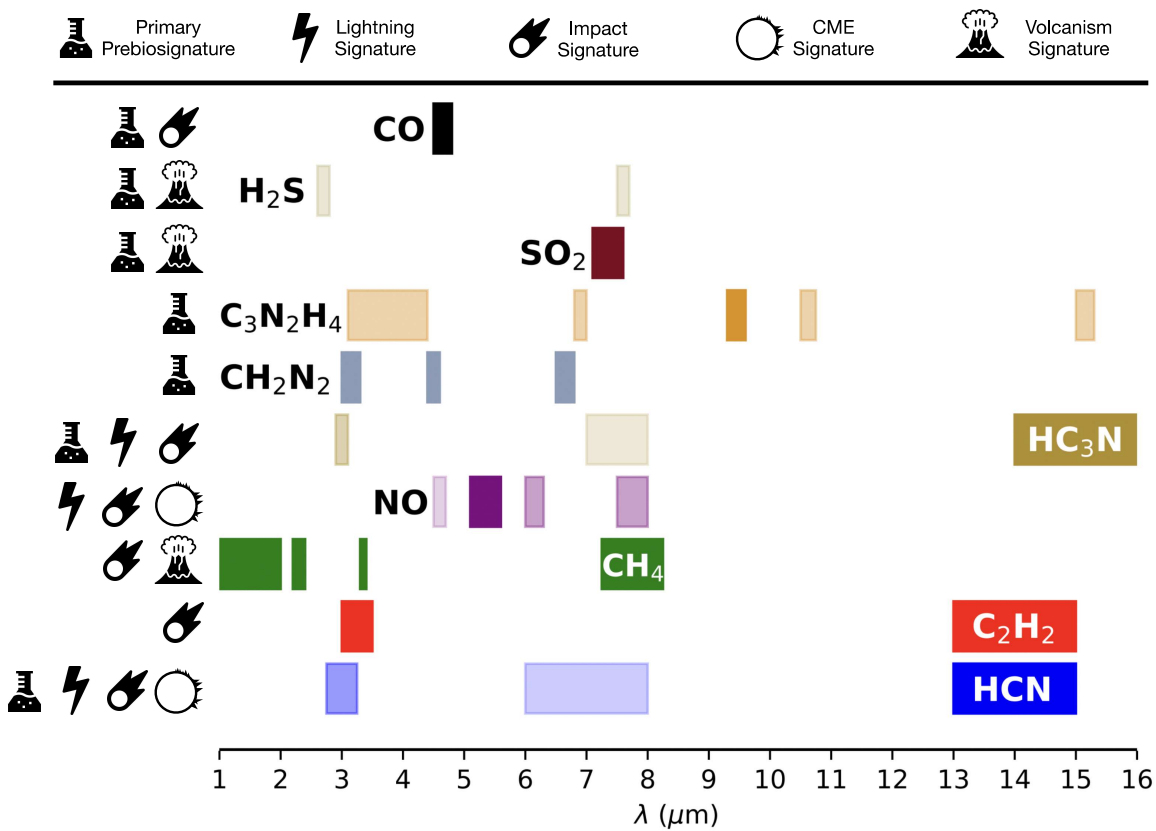}
    \caption{Reproduced from \citet{rimmer2021life}. Primary spectral features of a selection of prebiosignature molecules at JWST-probable wavelengths. The icons represent major physical and geochemical sources of the molecules, with primary prebiosignatures being directly associated with prebiotic chemistry. The species included are carbon monoxide (CO), hydrogen sulfide (\ch{H2S}), sulfur dioxide (\ch{SO2}), imidazole (\ch{C3N2H4}), cyanamide (\ch{CH2N2}), nitric oxide (\ch{NO}), methane (\ch{CH4}), acetylene (\ch{C2H2}), and hydrogen cyanide (\ch{HCN}). We include all of these except imidazole and cyanamide in our study. We also investigate ammonia and formaldehyde as potential prebiosignatures. Transparency indicates the strength of the features, with weaker features being more transparent. CME = Coronal Mass Ejection.}
    \label{fig:prebiosignatures}
\end{figure}

\hyperref[fig:prebiosignatures]{Figure 1}, from \cite{rimmer2021life}, details major prebiosignature molecules, their sources and their spectral features. We have chosen to focus our analysis on a selection of atmospheric species with wide prebiotic relevance, informed by the cyanosulfidic scenario \citep{patel2015common} and multiple other prebiotic pathways \citep[e.g.,][]{oro1961synthesis, ferris1968studies, miyakawa2002prebiotic,ferus2020one}: hydrogen cyanide (\ch{HCN}), sulfur dioxide (\ch{SO2}), hydrogen sulfide (\ch{H2S}), cyanoacetylene (\ch{HC3N}), carbon monoxide (\ch{CO}), methane (\ch{CH4}), acetylene (\ch{C2H2}), ammonia (\ch{NH3}), nitric oxide (\ch{NO}), and formaldehyde (\ch{CH2O}). The species were also chosen \sed{because they are relatively stable molecules that have measured infrared spectra. A brief summary of the relevance of these molecules to prebiotic chemistry follows below.}

\ch{HCN} is central to many prebiotic syntheses \citep[e.g.,][]{oro1961synthesis,sutherland2016origin}, \sed{including homologation to form adenine, one of the purine nucleotides used by life, and the triple bond between the carbon and nitrogen makes it a useful source of accessible chemical energy to form other prebiotically relevant carbon-containing molecules, such as simple sugars, pyrimidine ribonucleotides, and amino acids}. \sed{HCN was one of the main volatiles produced by Miller's experiment \citep{miller1953production}, and} can be produced by lightning, impacts, and photochemistry \citep{rimmer2019hydrogen}. \ch{HC3N} is also the primary feedstock of many pathways \citep[e.g.,][]{ferris1968studies,okamura2019one}, \sed{providing a chemical ``backbone'' for nucleotides and a pathway to forming a handful of amino acids. HCN} can be produced photochemically, as occurs on Titan \citep{clarke1997titan} and potentially on exoplanet GJ 1132b \citep{rimmer2021detectable}. Other scenarios utilise a somewhat reducing atmosphere of \ch{CO}, along with potentially \ch{CH4} or \ch{NH3}, as reagents for prebiotic experiments \citep[e.g.,][]{schlesinger1983prebiotic,miyakawa2002prebiotic}. These species can accumulate in an atmosphere due to volcanism from a reducing mantle \citep{liggins2022_jgr}, or from impacts \citep{zahnle2020creation}. The sulfur-bearing molecules \ch{H2S} and \ch{SO2} have been proposed as important catalysts in the cyanosulfidic \citep{patel2015common,xu2018photochemical} and carboxysulfitic \citep{liu2021prebiotic} chemistries, and both are products of volcanism, which itself may be an important component of the origin of life in the form of submarine \citep[e.g.][]{miller1988submarine} or surface \citep{rimmer2019origin} hydrothermal vents. Atmospheric \ch{C2H2} is another feature of impacts \citep{rimmer2019identifiable}, and large impacts can create transiently hydrogen-rich atmospheres \citep{zahnle2020creation,itcovitz2022reduced} well-suited for prebiotic synthesis \citep{benner2020did} in terrestrial planets. \sed{This is because impacts, and the reduced atmospheres that can result from giant impacts, provide a background chemical environment that can produce large quantities of \ce{HCN}, possibly \ce{HC3N}, and other prebiotically relevant compounds mentioned above.} \ch{CH2O} is another prebiotically relevant molecule \citep[e.g.][]{cleaves2008prebiotic}. \sed{It is the feedstock for the formose reaction \citep{butlerow1861formation}, a prominent prebiotic scheme that can lead to the formation of several simple sugars used by life \citep[e.g.][]{schwartz1993prebiotic}. In addition, formaldehyde reacts with cyanide to produce glycolonitrile, an intermediate that can lead to the production of glycolaldehyde and imidazolide intermediates on the way to RNA and DNA precursors \citep{patel2015common,xu2018photochemical,green2021illuminating}} and can also be formed from impacts \citep{ferus2019prebiotic}. \sed{Nitric Oxide (\ch{NO})} is an important indicator of lightning, stellar activity, and impacts \citep{mvondo2001production,airapetian2016prebiotic,heays2022nitrogen} \sed{and all of these processes can lead to the production of \ce{HCN} and other prebiotically relevant molecules listed above. It is therefore a potential tracer of physical processes that can lead to the production of life's building blocks.}

\subsection{Planetary Atmospheres}

The atmospheres of exoplanets are observable by transmission spectroscopy \citep[e.g.][]{charbonneau2002detection}, secondary eclipse spectroscopy \citep[e.g.][]{stevenson2010possible}, and direct imaging \citep[e.g.][]{konopacky2013detection}. Critical to our investigation is the ability to retrieve the abundances of particular molecules from the strength of their associated spectral features, for which transmission spectroscopy in the infrared is particularly well-suited; hence our focus on \textit{JWST} for detecting prebiosignatures. 

Already with the \textit{Hubble Space Telescope (HST)} we have detections of atmospheres for super-Earths \citep[e.g.][]{tsiaras2016detection} and even terrestrial planets like GJ 1132b \citep{southworth2017detection,swain2021detection} around M stars. Other \textit{HST} observations resulted in flat-spectra \citep[e.g.][]{berta2012flat}, including of GJ 1132b \citep{mugnai2021ares,libby2022featureless}. Such flat spectra are explainable by high clouds \citep[e.g.][]{kreidberg2014clouds}, high mean molecular weight atmospheres, or a complete lack of an atmosphere. \textit{JWST} offers a further leap in sensitivity, and operates at a favourable wavelength range for molecular absorption features in the near infrared (NIR) and mid infrared (MIR) between 1 and 10 $\mu$m. This offers an opportunity to explore habitable worlds, particularly around M stars \citep[e.g.,][]{morley2017observing,wunderlich2021detectability}. 

Transmission spectroscopy is limited to transiting exoplanets, which represent the majority of known exoplanets, but only a small fraction of all exoplanets. Direct imaging offers the opportunity to explore non-transiting exoplanets, and is better suited to FGK stars \citep[e.g.,][]{des2002remote,rugheimer2013spectral}. Future wide-aperture ground-based telescopes, and potential space missions such as the Near-Infrared/Optical/Ultraviolet telescope proposed by the US Astro2020 Decadal Survey in reflected starlight, and the \textit{Large Interferometer For Exoplanets (LIFE)} in thermal emission \citep{quanz2021large}, may offer the opportunity to directly explore the atmospheres of Earth-like planets around Sun-like stars and other non-transiting exoplanets. \sed{\textit{LIFE} has sufficient spatial resolution to also be suitable for M dwarfs \citep{quanz2022large}}. While both transmission spectroscopy and direct imaging will likely contribute substantially to our observations of exoplanet atmospheres in the future, this work will focus on transmission spectroscopy to play to the strengths of \textit{JWST}. This means targeting smaller M and K star systems due to the stronger signal of transits around small stars. This comes with both advantages and disadvantages. Habitable zone planets are especially abundant around M stars \citep{dressing2015occurrence}, but an M-star host also has considerable implications for the habitability and prebiotic chemistry of the planets \citep{scalo2007m,shields2016habitability,rimmer2018origin}.

Given how widely anticipated the launch of \textit{JWST} has been, there have been many studies into its ability to explore exoplanet atmospheres \citep[e.g.,][]{morley2017observing,batalha2018strategies,wunderlich2019detectability}. \citet{morley2017observing} test the detectability of Venus-, Titan- and Earth-like atmospheres of known terrestrial exoplanets at the 5$\sigma$ level, finding it possible in most cases but expensive in terms of observation times. Notably, seven of the nine planets studied require fewer than 20 transits for the detection of a high mean molecular weight atmosphere. \citet{batalha2018strategies} calculate that the predominant atmospheric gas is detectable by 10 transits for temperate terrestrial planets around M stars. Of particular relevance to our analysis is the ability to detect trace molecules, and much work has been done on this for biosignatures \citep[e.g.,][]{krissansen2018detectability,schwieterman2018exoplanet,lustig2019detectability}. By far the most promising atmospheres for detection of biosignatures are hydrogen-rich atmospheres \citep{seager2013biomass,seager2013biosignature,schwieterman2018exoplanet,madhusudhan2021habitability} due to the low mean molecular weight resulting in a puffy atmosphere. \citet{swain2021detection} claim a prebiosignature molecule (HCN) in such an atmosphere around GJ 1132b, although this is disputed \citep{mugnai2021ares,libby2022featureless}. \citet{rimmer2021detectable} use this claim along with photochemical models to predict a \textit{JWST}-detectable abundance of another prebiosignature (\ch{HC3N}) in this planet's atmosphere.

Because of the advantages low mean molecular weight atmospheres present for making molecular detections, we focus on these types of atmospheres on model planets around M stars.  We explore the limits that mean molecular weight, star radius, and observation strategy place on the detectability of prebiosignatures.

\subsection{Outline of Work}

In this work, we assess the detectability of our selected prebiosignatures in a collection of physically-motivated background atmospheres using \sed{Bayesian detection tests on simulated} \textit{JWST} transmission spectra. We hope to lay the groundwork for the search for prebiotic environments in the age of \textit{JWST}, by compiling a list of detection thresholds for each prebiosignature molecule.  We also present {\tt TriArc}, a versatile and flexible Bayesian detection test tool, for calculating the detection thresholds for atmospheric species in a prescribed background atmosphere.

We describe in \hyperref[sec:Method]{Section 2} the creation of a detection threshold pipeline to calculate the minimum abundance of a prebiosignature required for a 3$\sigma$ detection using radiative transfer modelling, synthetic JWST noise, and a fully Bayesian detection test targeted on individual spectral bands. Using a multi-band approach the abundance of multiple species can be constrained simultaneously. We present the results of our work in \hyperref[sec:Results]{Section 3}, including model transmission spectra, a library of spectral features of prebiosignatures, and the grid of detection thresholds. In \hyperref[sec:Discussion]{Section 4} we discuss the impacts of planetary and observational properties on the detection thresholds, and briefly consider the consequences of the thresholds for prebiotic chemistry and other planetary processes. In \hyperref[sec:Conclusion]{Section 5} we summarise our conclusions and look to the future of observing prebiosignatures.

\section{Method}
\label{sec:Method}

To investigate the detectability of prebiosignatures, we \sed{have developed a novel detection test tool {\tt TriArc}, leveraging the forward modelling of the open-source package {\tt petitRADTRANS (pRT)} \citep{molliere2019petitradtrans}. We use {\tt pRT} to simulate transmission spectra (\hyperref[subsec:petitRADTRANS]{Section 2.1}), add synthetic \textit{JWST} noise (\hyperref[subsec:PandExo]{Section 2.2}), and then perform Bayesian inference against an array of test spectra varying a single parameter, using the statistical framework of {\tt TriArc} to obtain a significance of detection and hence a detection threshold (\hyperref[subsec:TriArc]{Section 2.3}). 
}

\subsection{Atmospheric Models}
\label{subsec:petitRADTRANS}

For the detection tests and atmospheric retrieval, we need to create a forward model of a background atmosphere with a prescribed abundance of a prebiosignature molecule.  The transmission spectra of this reference atmosphere is then compared to a set of hypothesis transmission spectra for atmospheres containing various amounts of the prebiosignature. To synthesise these transmission spectra, we calculate the 1D radiative transfer using the correlated-\textit{k} approximation mode of the {\tt petitRADTRANS}\footnote{https://petitradtrans.readthedocs.io/en/latest/} {\tt (pRT)} package \citep{molliere2019petitradtrans}. For each atmosphere, we use 100 layers of atmosphere equally distributed in log-space from 1 $\mu$bar down to the surface pressure, and an isothermal temperature profile. We use rebinned opacities at a spectral resolution of R=100. We consider line opacities from the following species: \ch{H2O} and \ch{CO} \citep{rothman2010hitemp}, \ch{CO2} \citep{yurchenko2020exomol}, \ch{CH4} \citep{yurchenko2017hybrid}, \ch{HCN} \citep{barber2014exomol}, \ch{NH3} \citep{coles2019exomol}, \ch{H2S} \citep{azzam2016exomol}, and \ch{C2H2} \citep{chubb2020exomol}; Rayleigh opacities from \ch{H2}, \ch{He}, \ch{N2}, \ch{CH4} and \ch{H2O}; and collision-induced-absorption (CIA) from \ch{H2-H2}, \ch{H2-He}, and \ch{N2-N2}, all included within {\tt pRT}. We also include line opacities from \ch{NO} \citep{wong2017exomol}, \ch{SO2} \citep{underwood2016exomol}, and \ch{CH2O} \citep{al2015exomol} from the ExoMol database \citep{tennyson2016exomol}, and \ch{HC3N} from \citet{rimmer2021detectable}. We also assume vertically-uniform mixing ratios of the atmospheric species.

\subsubsection{Background Atmospheres}

{\tt TriArc} \sed{requires a ``background atmosphere'' composition, which is held constant during the prebiosignature detection threshold calculation. This includes both continuum opacity sources and line opacities, with abundances dictated by chemical equilibrium or kinetic modelling.} It is therefore important for us to present a set of reasonable background atmospheres in which we can test for prebiosignatures. Our own solar system hosts a diverse set of planetary atmospheres, each determined by their planets' distinct geological evolution \citep[e.g., see][]{pierrehumbert2010principles}. Venus, Earth, and Titan are all rocky bodies with dense atmospheres, and \citet{morley2017observing} use them as models for the detectability of terrestrial atmospheres. However, the atmospheres of Earth and Venus are insufficiently reducing to generate significant RNA precursor prebiosignatures like HCN \citep{benner2020did}. Titan on the other hand is an ideal environment for many prebiosignatures, with species like HCN, \ch{C2H2} and \ch{HC3N} present in the atmosphere \citep{sagan1992titan,khanna2005condensed}. However, all of these planets have atmospheres poorly suited to detection due to their high mean molecular weight. For an ideal chance of atmospheric characterisation, we seek planets with large scale heights (and hence low mean molecular weights, and high temperatures) orbiting small stars (M and late K dwarfs). This can be clearly identified by writing out the effective transit depth, $R_t = R_{pl} + z_{atm}$, where $R_{pl}$ is the radius of planet's surface, and $z_{atm}$ is some characteristic height of the atmosphere at that wavelength (and some function of the scale height, $z_{atm} \ll R_{pl}$). The transmission spectrum is given by the transit depth over the stellar radius $R_*$ squared:

\begin{equation}
    \left (\frac {R_t}{R_*} \right)^2 \approx \left (\frac {R_{pl}}{R_*} \right)^2 + \frac{2 R_{pl} z_{atm}}{R_*^2}.
\end{equation}

As the first term is independent of wavelength, the signal of the transmission spectrum (the wavelength dependent second term) is seen to be proportional to the atmospheric height $z_{atm}$, which in turn can be related to the atmospheric scale height in hydrostatic equilibrium

\begin{equation}
    H = \frac{k_B T}{\mu g m_H},
\end{equation}

\noindent where $T$ is the temperature, $\mu$ is the mean molecular weight, $g$ is the gravitational acceleration, $k_B$ is Boltzmann's constant, and $m_H$ is the mass of a hydrogen atom.

Therefore, instead of considering solar system planetary atmospheres, we primarily select planets with atmospheres rich in hydrogen (\ch{H2}) and helium (\ch{He}) so that they have a reduced mean molecular weight. This has already been considered as a promising angle for biosignature analysis \citep{seager2013biosignature}. We are accustomed to giant planets possessing \ch{H2}-dominated atmospheres, but do not see \ch{H2} in the atmospheres of rocky solar systems planets due to the sensitivity of \ch{H2} to hydrodynamic, hydrostatic and non-thermal escape \citep[e.g.,][]{kasting1983loss,lammer2008atmospheric,zahnle2017cosmic}. In order to justify the existence of \ch{H2} in our atmospheric models, we turn to a small number of mechanisms. More massive planets (super-Earths and mini-Neptunes) are capable of retaining primordial \ch{H2}-rich envelopes \citep{fortney2007planetary,ginzburg2016super}, and could potentially include a liquid water surface \citep{madhusudhan2020interior}. Terrestrial planets may be able to retain their primordial atmospheres at greater distances from their star, and remain habitable due to the greenhouse effect of \ch{H2} \citep{pierrehumbert2011hydrogen}. Volcanic outgassing could also maintain \ch{H2} in the atmosphere \citep{tian2005hydrogen,ramirez2017volcanic,liggins2020can}, and even result in an \ch{H2}-dominated atmosphere with a sufficiently reduced mantle \citep{swain2021detection}. \citet{zahnle2020creation} demonstrate that accretion of reducing iron in the aftermath of an impact can result in a transiently hydrogen-dominated atmosphere.

With the above considerations, we choose the following physically-motivated background atmospheres: a Hycean world (ocean planet with hydrogen atmosphere), an ultrareduced volcanic world (active volcanic planet with hydrogen- and nitrogen-rich outgassing), and a post-impact world (planet in the aftermath of a collision with another planetary body resulting in evaporation of the oceans and reduction of atmospheric species by metals from the impacting body) at two different times after the collision. We also include a super-Earth planet similar to that considered by \citet{seager2013biosignature} to represent a theoretical class of rocky super-Earths with thin hydrogen envelopes that could be formed by outgassing after oxidation of metallic iron by accreted water \citep{elkins2008ranges}.

The Hycean world is drawn from the planetary properties and atmospheric composition used by \citet{madhusudhan2021habitability} for the planet K2-18b, constrained by the observations of \citet{benneke2019water}. As a potentially habitable planet, K2-18b is a suitable candidate for both prebiosignature and biosignature analysis. The ultra-reduced volcanic world is based on the \sed{hydrogen-rich outgassed atmosphere modelled by \citet{swain2021detection} to explain the detection of HCN in GJ 1132b. The detection is refuted by \citet{mugnai2021ares} and \citet{libby2022featureless}, but the atmosphere considered is nonetheless prebiotically relevant: photochemical modelling of this ultrareduced volcanic atmosphere predict an abundance of over 1 ppm of the key prebiosignature \ch{HC3N} at 10 mbar \citep{rimmer2021detectable} in addition to HCN. Such an atmosphere} is therefore a logical candidate for a more comprehensive analysis. The post-impact planets are based on the modelling of the effects of impacts on the Hadean Earth by \citet{zahnle2020creation}, considering the atmospheres present at 0.1 Myr and 10 Myr after the impact respectively. Impacts, which lead to substantial amounts of reducing iron (i.e. from the core of the impactor), are proposed as a method for an Earth-like planet to achieve a transient reducing atmosphere and hence a suitable environment for the origin of life \citep{benner2020did}. The post-impact atmospheres we present are end-member cases, being the most reducing possible after the impacts of \citet{zahnle2020creation}. A variety of processes during the impact can lead to less reducing post-impact environments \citep{itcovitz2022reduced}. The super-Earth uses an atmospheric composition derived from the photochemical results of \citet{hu2012photochemistry}.

\begin{deluxetable}{ccccccc}[ht]
\label{tab:planets}
\tablehead{\colhead{Model Planet} & \colhead{Radius} & \colhead{Gravity} & \colhead{Temperature} & \colhead{Surface Pressure} & \colhead{MMW} & \colhead{Scale Height} \\ 
\colhead{} & \colhead{($\mathrm{R_{Earth}}$)} & \colhead{($\mathrm{ms^{-2}}$)} & \colhead{(K)} & \colhead{(bar)} & \colhead{} & \colhead{(km)} }
\tablecaption{Planetary Properties of Model Exoplanets for Prebiosignature Detection Threshold Analysis}
\startdata
Super-Earth & 1.70 & 13.93 & 290 & 1.0 & 4.60 & 38\\
Hycean & 2.51 & 13.56 & 300 & 100 & 2.35 & 78\\
Ultrareduced Volcanic & 1.20 & 10.89 & 480 & 1.0 & 4.51 & 81\\
100 kyr Post-Impact & 1.00 & 9.81 & 435 & 55 & 2.38 & 155\\
10 Myr Post-Impact & 1.00 & 9.81 & 424 & 45 & 2.25 & 160\\
TRAPPIST-1e/Early Earth & 0.91 & 9.12 & 246 & 1.0 & 29.6 & 7.6\\
\enddata
\end{deluxetable}

\begin{deluxetable}{cccccccccc}[!ht]
\label{tab:atmospheres}
\tablehead{\colhead{Model Planet} & \colhead{\ch{H2}} & \colhead{\ch{He}} & \colhead{\ch{N2}} & \colhead{\ch{CH4}} & \colhead{\ch{CO}} & \colhead{\ch{CO2}} & \colhead{\ch{H2O}} & \colhead{\ch{HCN}} & \colhead{\ch{NH3}}  }
\tablecaption{Atmospheric Mixing Ratios of Model Exoplanets for Prebiosignature Detection Threshold Analysis}
\startdata
Super-Earth & 90\% & 0.0 & 10\% & 0.0 & 0.0 & 1e-4 & 1e-5 & 0.0 & 0.0\\
Hycean & 90\% & 9\% & 0.0 & 5e-4 & 0.0 & 0.0 & 1\% & 0.0 & 1e-4\\
Ultrareduced Volcanic & 90\% & 0.1\% & 8.9\% & 0.3\% & 0.3\% & 0.0 & 2e-5 & 0.3\% & 0.0\\
100 kyr Post-Impact & 98\% & 0.0 & 0.5\% & 1.76\% & 0.0 & 0.0 & 1e-7 & 0.0 & 0.0\\
10 Myr Post-Impact & 99\% & 0.0 & 0.41\% & 0.0 & 0.575\% & 0.0 & 1e-7 & 0.0 & 0.0\\
TRAPPIST-1e/Early Earth & 0.0 & 0.0 & 90\% & 2e-6 & 0.0 & 10\% & 1e-6 & 0.0 & 0.0\\
\enddata
\end{deluxetable}

\begin{figure}
    \centering
    \includegraphics[width=0.45\textwidth]{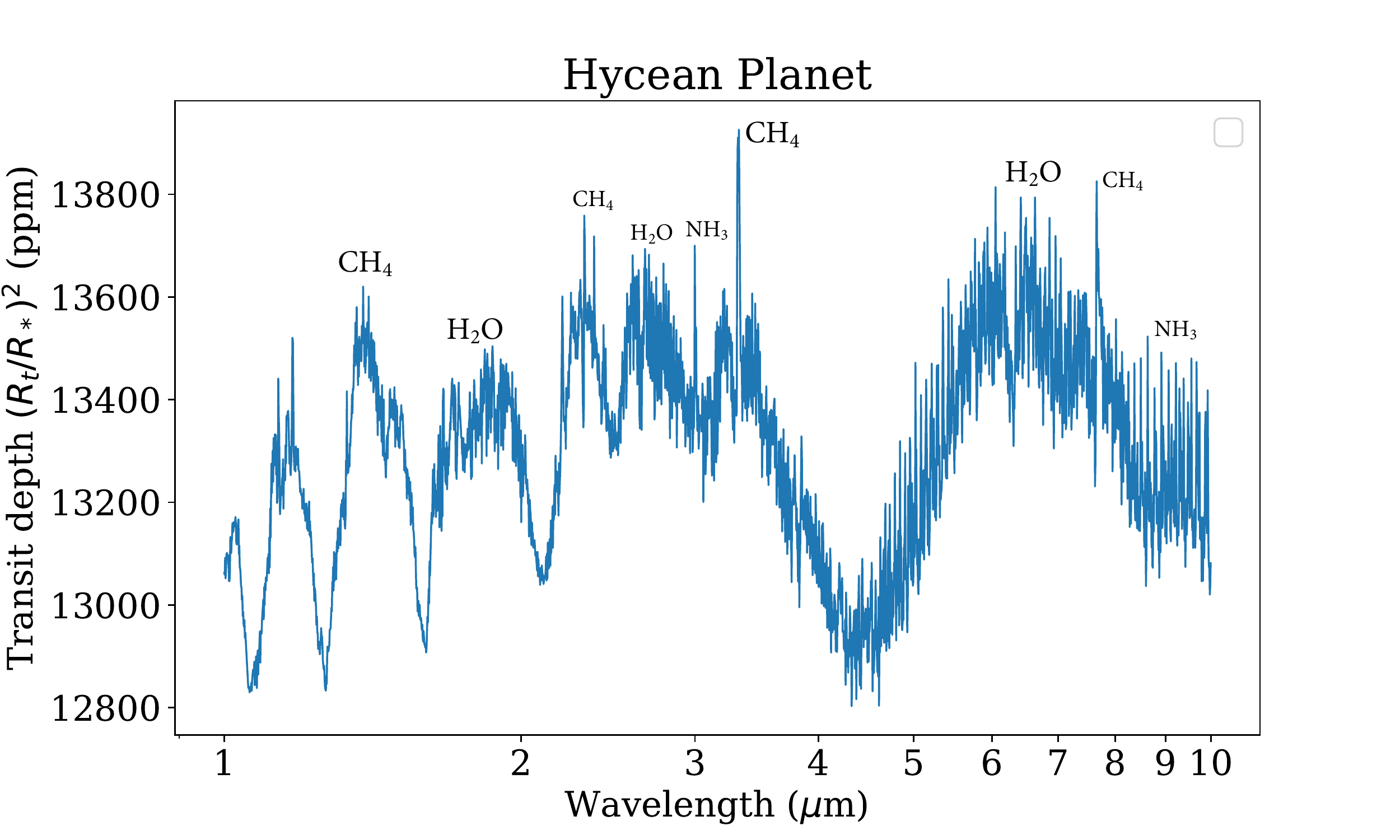}
    \includegraphics[width=0.45\textwidth]{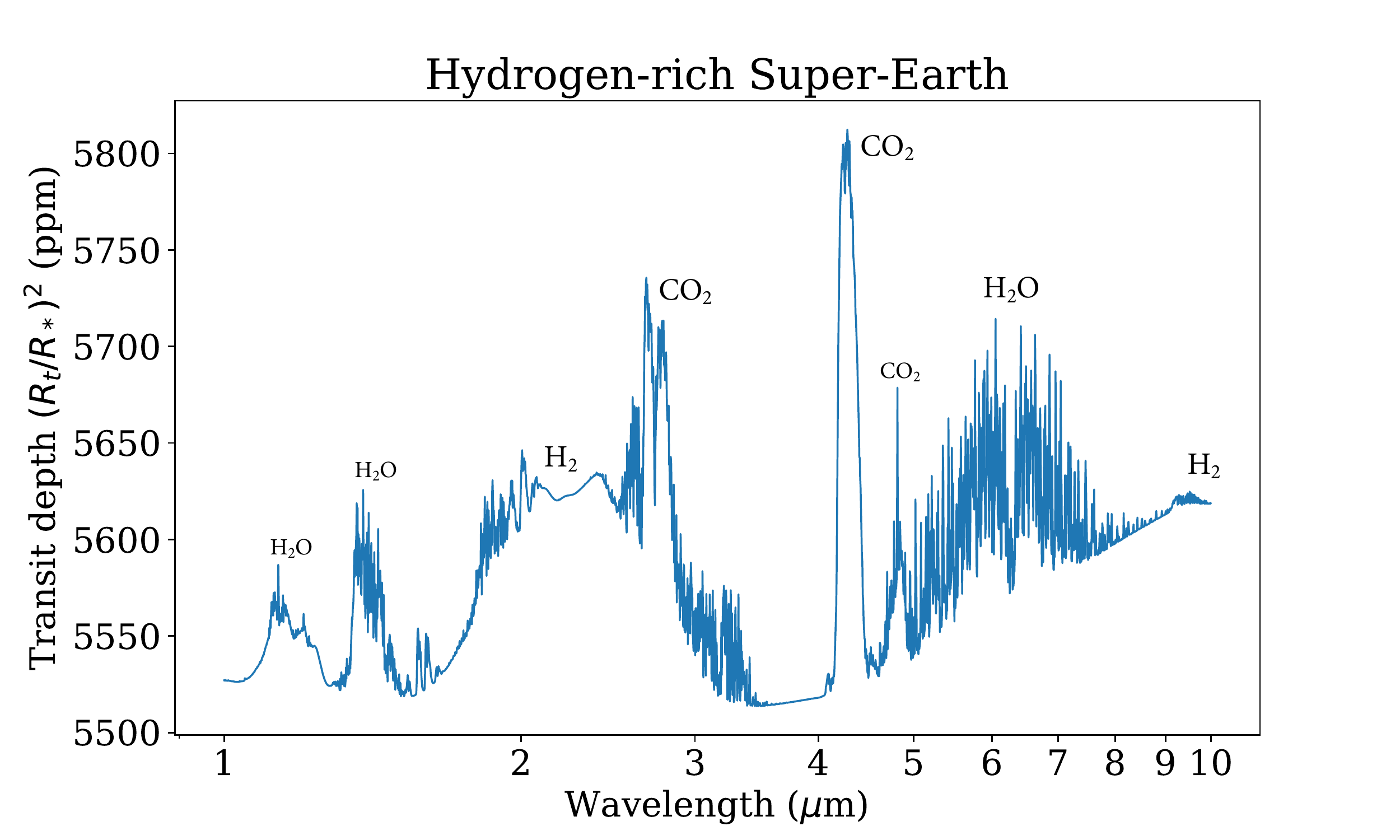}
    \includegraphics[width=0.45\textwidth]{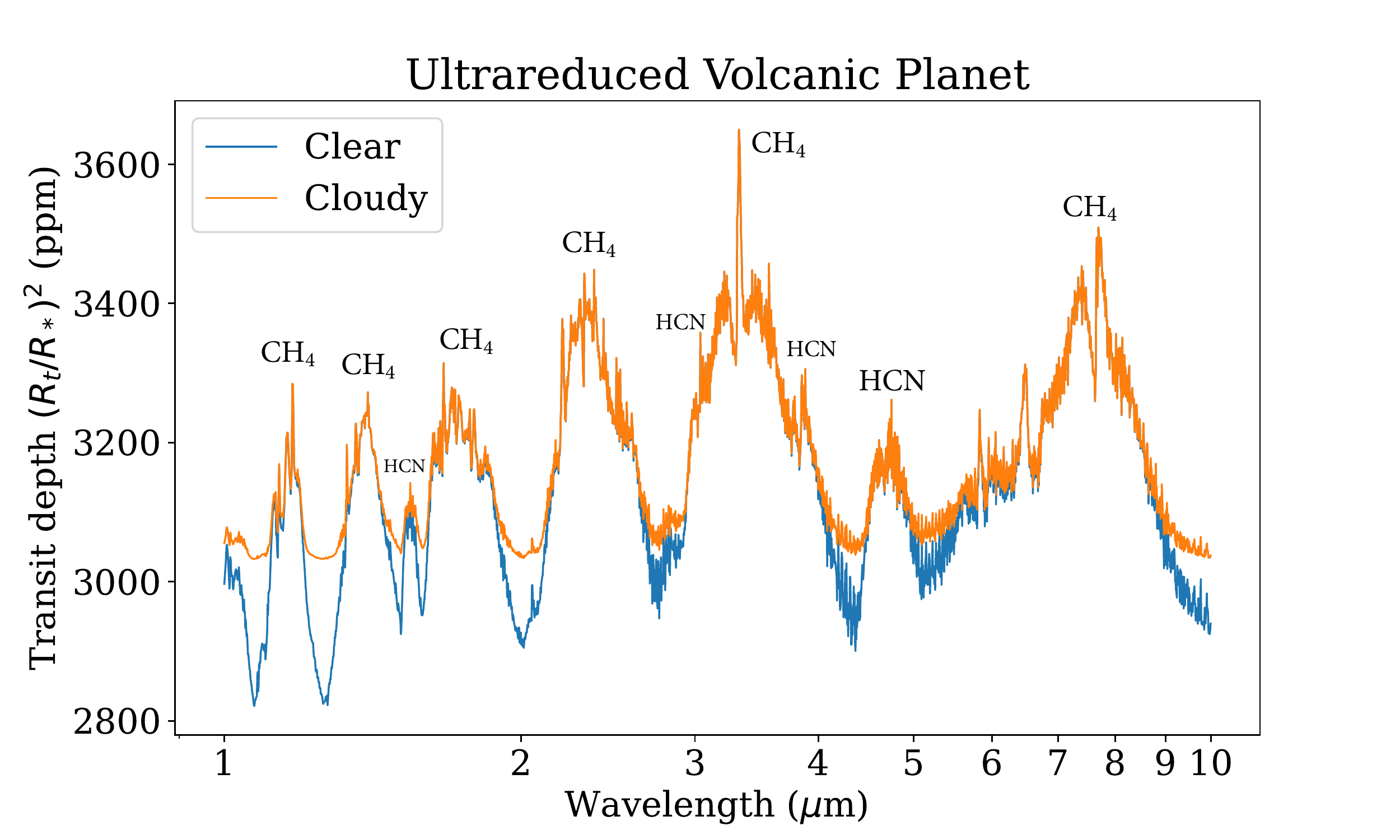}
    \includegraphics[width=0.45\textwidth]{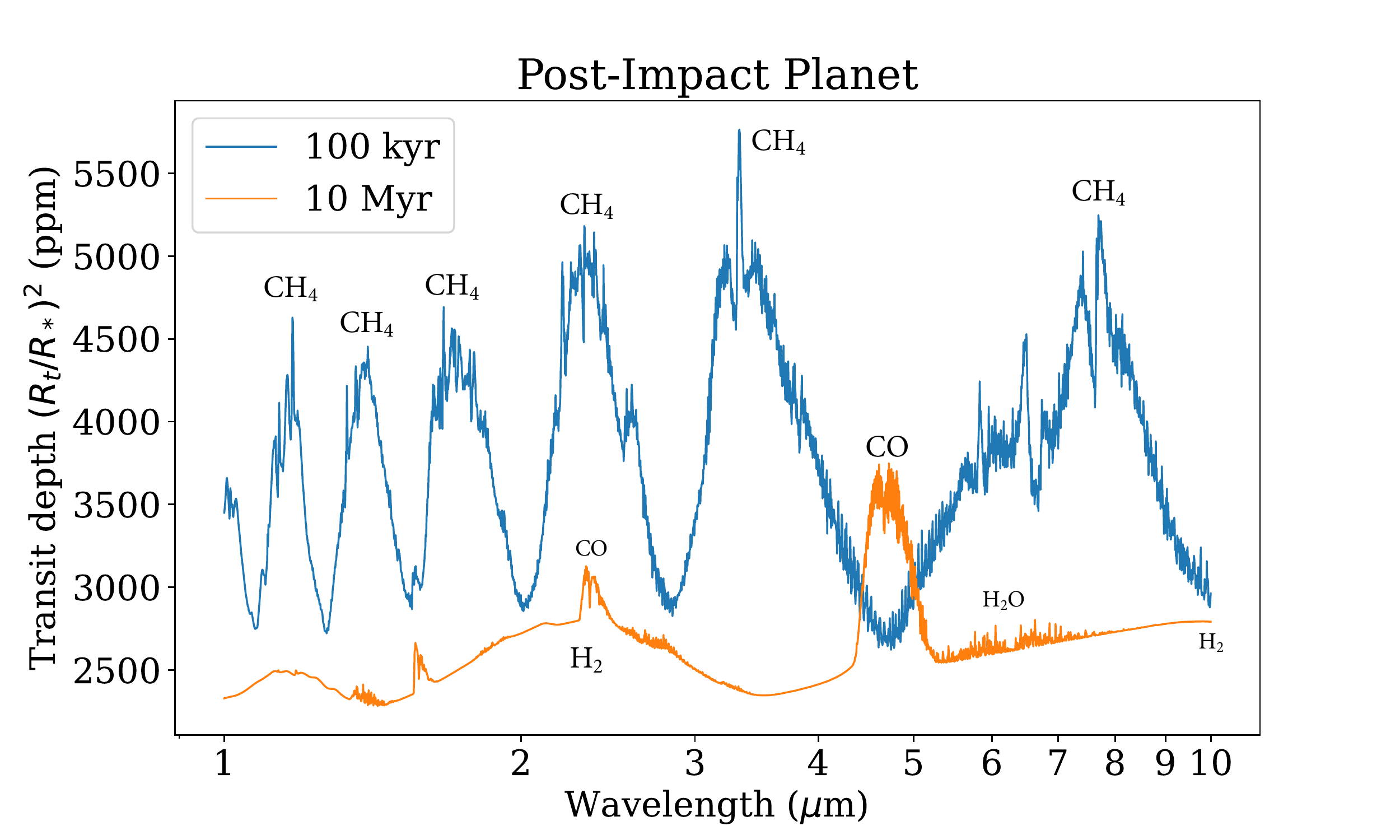}
    \includegraphics[width=0.45\textwidth]{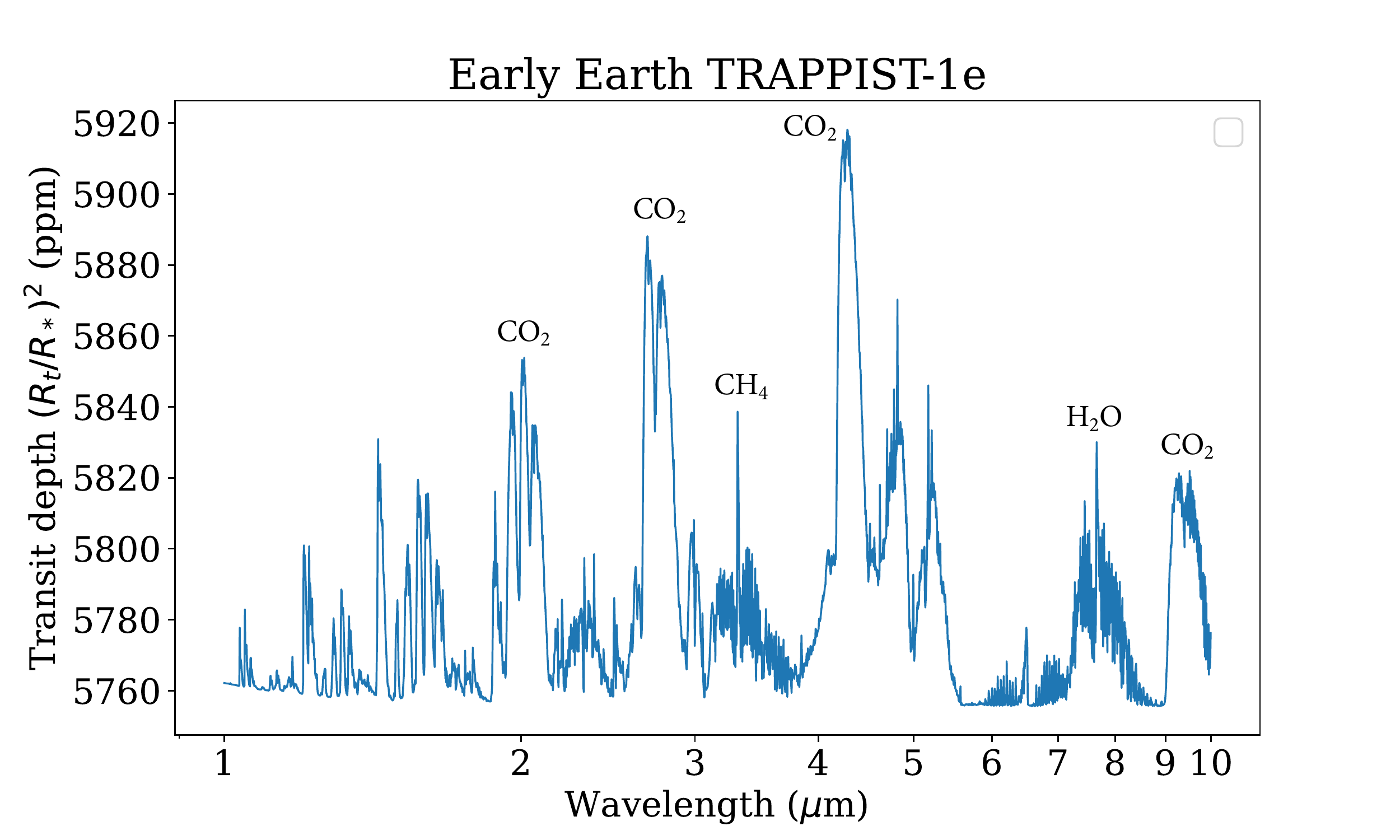}
    
    \caption{ Transmission spectra of model exoplanets, calculated using {\tt petitRADTRANS}. Top left: Hycean planet, with planetary and atmospheric properties chosen to match those described by \cite{madhusudhan2021habitability} for K2-18b. Top right: Super-Earth with thin hydrogen envelope, with planetary properties chosen to match those of HD 88512b, and atmospheric properties based on the modelling of \citet{hu2012photochemistry}. Middle left: Ultrareduced Volcanic planet, both with (blue) and without (orange) a grey cloud layer at 10 mbar, with planetary and atmospheric properties chosen to match those described by \cite{swain2021detection} for GJ 1132b. Middle right: Post-Impact planet transmission spectra 100 kyr (blue) and 10 Myr (orange) after the impact, with atmospheric properties chosen to match those described by \cite{zahnle2020creation} for a Hadean Earth. Bottom: Model TRAPPIST-1e with an atmosphere to emulate an early epoch of Earth history \citep{kaltenegger2007spectral}.}
    \label{fig:spectra}
\end{figure}

All the model atmospheres are assumed to be isothermal with vertically uniform mixing ratios, and free from clouds and hazes. The pressure-temperature profile does not have a significant impact in transmission spectroscopy \citep{morley2017observing}, but may be overly simplified and introduce bias into the retrieval \citep{rocchetto2016exploring}. The physical properties of these model planets are displayed in \hyperref[tab:planets]{Table 1}, and their atmospheric compositions are presented in \hyperref[tab:atmospheres]{Table 2}. \sed{The transmission spectra of these background atmospheres are presented in \hyperref[fig:spectra]{Figure 2}.}

We do not want to completely preclude high mean molecular weight atmospheres from our analysis, particularly as study of the early Earth is important to the understanding of the origin of life, and Earth has possessed a high mean molecular weight secondary atmosphere for most of its history. Furthermore, one of the most promising systems for exoplanetary analysis is TRAPPIST-1 \citep{gillon2016temperate}, a planetary system with seven terrestrial planets, of which some are in the habitable zone \citep[e.g.,][]{barstow2016habitable,lustig2019detectability}. Observations suggest that these planets do not possess hydrogen-rich atmospheres \citep{de2018atmospheric}. Therefore we include an additional model exoplanet, with planetary properties based on TRAPPIST-1e and the atmospheric composition based on a potential early epoch of Earth history \citep[see][]{kaltenegger2007spectral,rugheimer2015uv}.

For consistency, we model these planets as orbiting a benchmark M4V star with a radius of 0.21 $R_{\sun}$ (based on GJ 1132) in our initial analysis (\hyperref[subsec:Thresholds]{Section 3.1}). \sed{In a subsequent analysis (\hyperref[subsec:observational]{Section 3.4}) we use noise and spectra derived from the Hycean planet orbiting alternative M dwarf targets with different radii and magnitudes:} K2-3 and K2-18 \citep{hardegree2020scaling}, LTT 1445 A \citep{winters2019three}, and TRAPPIST-1 \citep{lienhard2020global}\sed{, to test the impact of varying these parameters on the detection thresholds. TRAPPIST-1 is modelled with its own spectral type in our analysis of a high mean molecular weight atmosphere for TRAPPIST-1e.}

\subsection{Synthetic \textit{JWST} Data}
\label{subsec:PandExo}

In order to analyse the ability of JWST to detect prebiosignatures, we use the {\tt PandExo}\footnote{https://natashabatalha.github.io/PandExo/} package \citep{batalha2017pandexo} to simulate realistic noise. For our consistent noise profile to test the impact planetary properties we devise a reasonable observation regime, simulating an M4V star based on GJ 1132 to acquire noise data, as \cite{morley2017observing} find GJ 1132b to be a good candidate for transmission spectroscopy. {\tt PandExo} uses the PHOENIX model library \citep{husser2013new} to simulate observations from stellar inputs.

\begin{figure}
    \centering
    \label{fig:instruments}
    \includegraphics[width=0.7\textwidth]{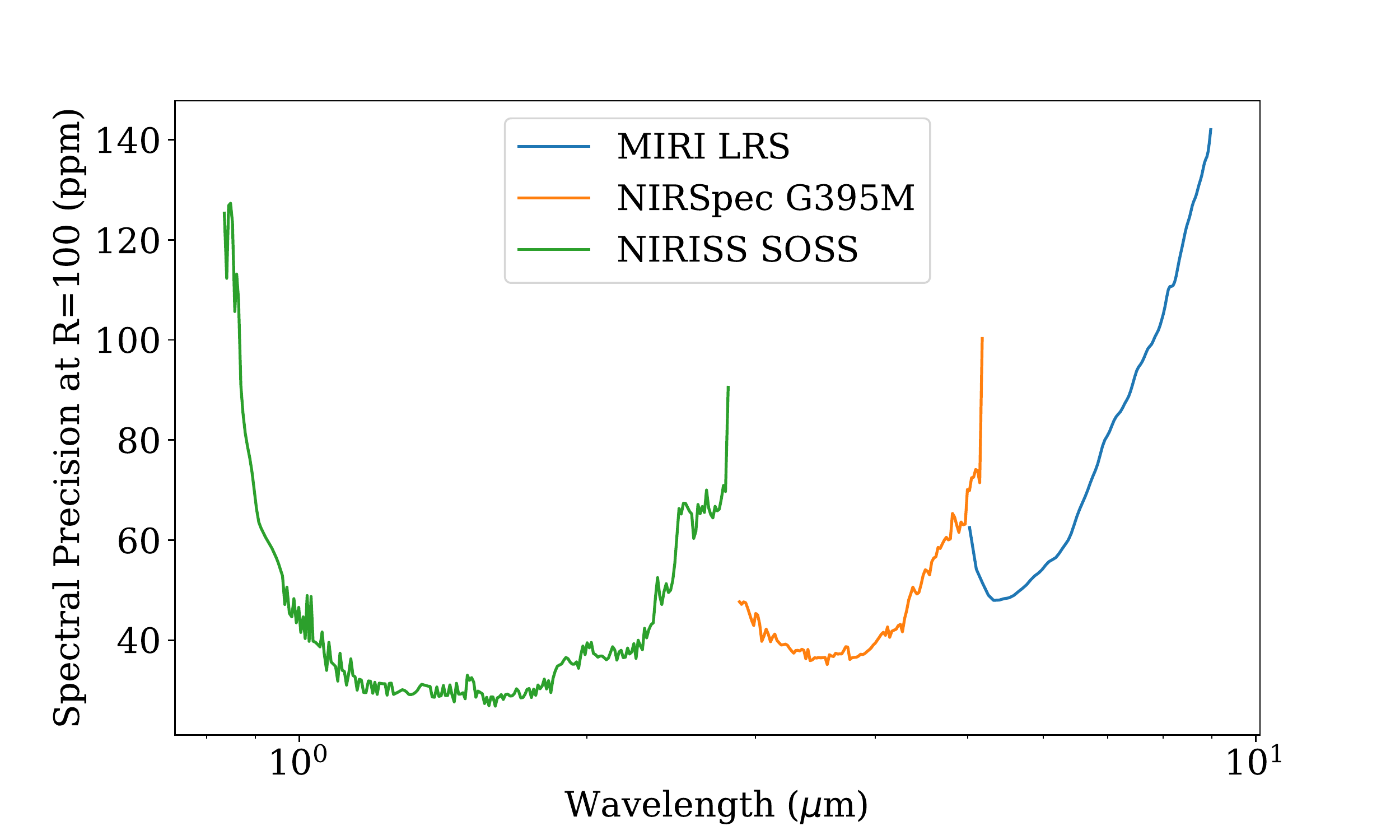}
    \caption{Spectral precision of JWST instruments as a function of wavelength, at resolution R=100 when observing GJ 1132b for 6 hours per instrument, including 3 transits, at 80\% full well saturation, simulated using {\tt PandExo}.}
\end{figure}

We simulate at a spectral resolution of $R=100$, using six \sed{total} hours of observation per instrument, including three transits \sed{(48 minutes each)} and a three hour baseline, at 80\% full well saturation. We find that using the following instruments to explore the entire wavelength range for 1-10 $\mathrm{ \mu m}$ is most effective: \sed{NIRISS SOSS ($0.8-2.9, \mathrm{ \mu m}$)}, NIRSpec G395M ($2.9-5 \, \mathrm{ \mu m}$) and MIRI LRS ($5-10 \, \mathrm{ \mu m}$), giving a good noise of 25-80 ppm (see \hyperref[fig:instruments]{Figure 3}). \sed{Although its wide spectral baseline makes it attractive for atmospheric characterization, NIRSpec Prism saturates for GJ 1132 and is near or below the saturation limit for many of the other terrestrial planet systems we may consider \citep{jakobsen2022near}.} The NIRISS SOSS/NIRSpec G395M combination contains the best information content for characterizing exoplanets orbiting bright stars \citep{batalha2017information}.\sed{We use the medium-resolution grism G395M over the high-resolution grism G395H, as the latter possesses a gap in its spectral range between 3.7--3.8 $\mu$m which contains relevant spectral information for prebiosignatures.}

The approach we take in implementing JWST noise into {\tt TriArc} follows that of \cite{madhusudhan2021habitability}. Using {\tt PandExo} we calculate the sensitivity of the instrument as a function of wavelength for the different instruments, and add a corresponding amount of gaussian noise to our synthetic transmission spectra before feeding them into the detection test. We also use {\tt exo-k} \citep{leconte2021spectral} to re-bin the opacities used in the atmospheric models into the desired spectral resolution of R=100.

We compute a variety of noise profiles to test the sensitivity of the detection thresholds to observational parameters (\hyperref[subsec:observational]{Section 3.4}). The process is the same as described above, except we vary the number of transits and the baseline observation time, which we state in each case.

\vspace{5mm}

\subsection{Detection Tests}
\label{subsec:TriArc}

In order to estimate the significance of detection for any particular atmosphere and noise profile, we perform a Baeysian detection test. We have developed {\tt TriArc} for that purpose, using Bayesian inference to retrieve a single parameter (the abundance of a single species) with all other parameters fixed as delta priors. As only a single parameter is retrieved (\sed{as opposed to many parameters simultaneously in a full retrieval}), the need for more involved sampling approaches is eliminated, greatly improving the speed of our analysis. This allows a large number of detection tests to be performed, which is important for finding the location of the detectability threshold. \sed{A realistic retrieval of an observed transmission spectrum would need to fit for every unknown parameter, including the planetary characteristics and atmospheric composition, a much more computational demanding task. Therefore the detection thresholds that we obtain from {\tt TriArc} are an order-of-magnitude estimate of what is possible with \textit{JWST} observation. To verify our detection thresholds, we benchmark our detection tests against full retrievals performed using the retrieval package of {\tt pRT}.}

The hypotheses to test with Bayes' theorem are represented by a set of test spectra with different abundances of the species being retrieved. The distribution of these hypotheses is given by a Jeffreys prior with uniform probability in log-space, varying the mass fraction from $10^{-11}$ to $1$. The `evidence' is a noisy model spectrum with a prescribed abundance of the retrieved species. All other atmospheric parameters not being retrieved are fixed as delta priors. In order to retrieve the abundance of the species, i.e. assign a likelihood to each hypothesis, the goodness of fit between the noisy model spectrum and each of the test spectra is computed within the wavelength range of the spectral feature of the species being retrieved. The goodness of fit is measured using a Gaussian radial basis likelihood function over a set of data points in wavelength space $\lambda$, where $E_\lambda$ is the model spectrum data point, $H_{i,\lambda}$ is the $i$th test spectrum data point, and $\sigma_\lambda$ is the model noise:

\begin{equation}
    P(E \mid H_i) = \prod_{\lambda} ({2 \pi \sigma_{\lambda}^2})^{-\frac{1}{2}} \exp{\left(-\frac{(H_{i,\lambda} - E_\lambda)^2}{2\sigma_{\lambda}^2}\right)} .
\end{equation}

\noindent This probability of the evidence given the hypothesis $P(E \mid H_i)$ is then converted to a posterior probability distribution function (PDF), $P(H_i \mid E)$, using Bayes' theorem and the aforementioned Jeffreys prior $P(H_i)$:

\begin{equation}
    P(H_i \mid E) = \frac{P(E \mid H_i) P(H_i)}{\sum_i P(E \mid H_i) P(H_i)}. 
\end{equation}

\begin{figure}
    \centering
    \includegraphics[width=0.45\textwidth]{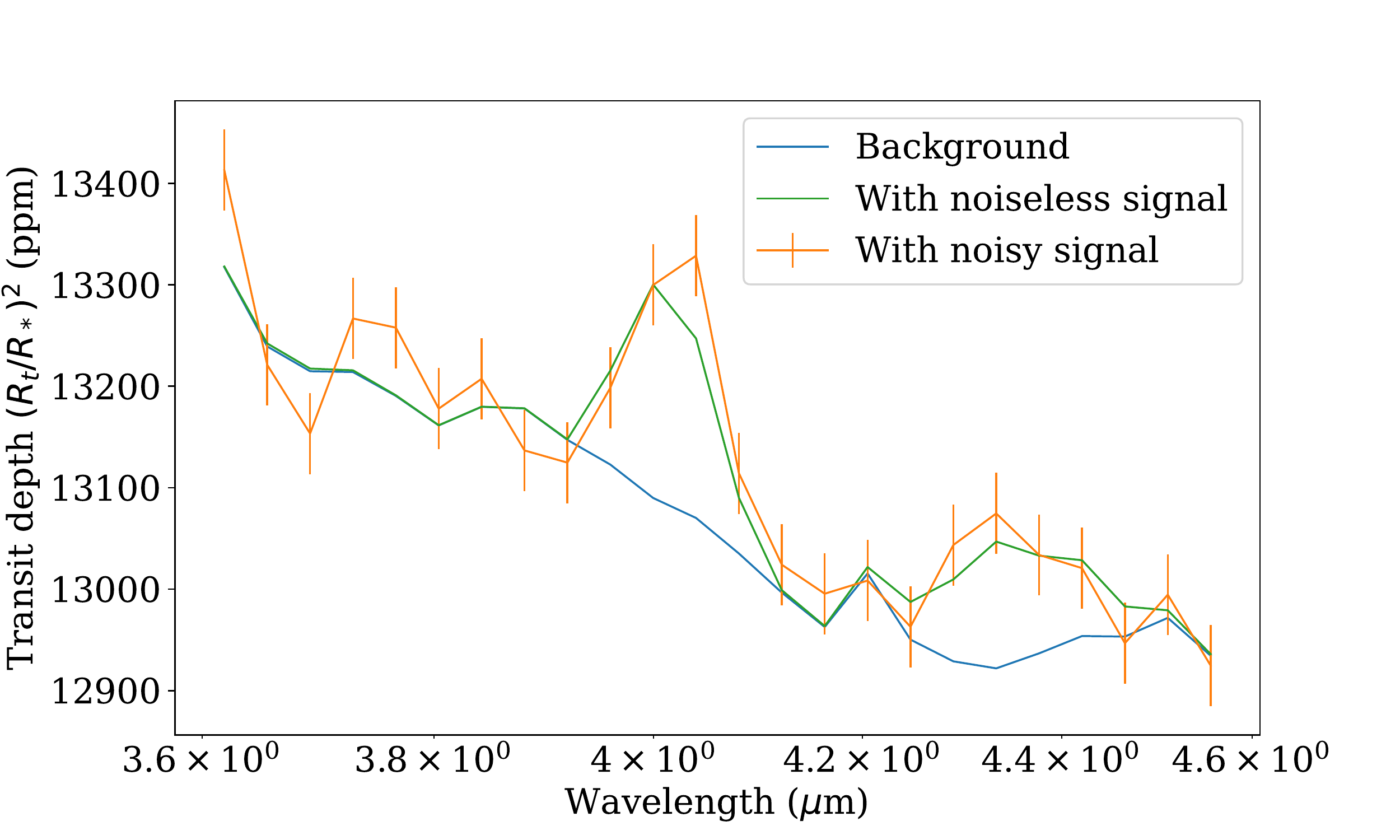}
    \includegraphics[width=0.45\textwidth]{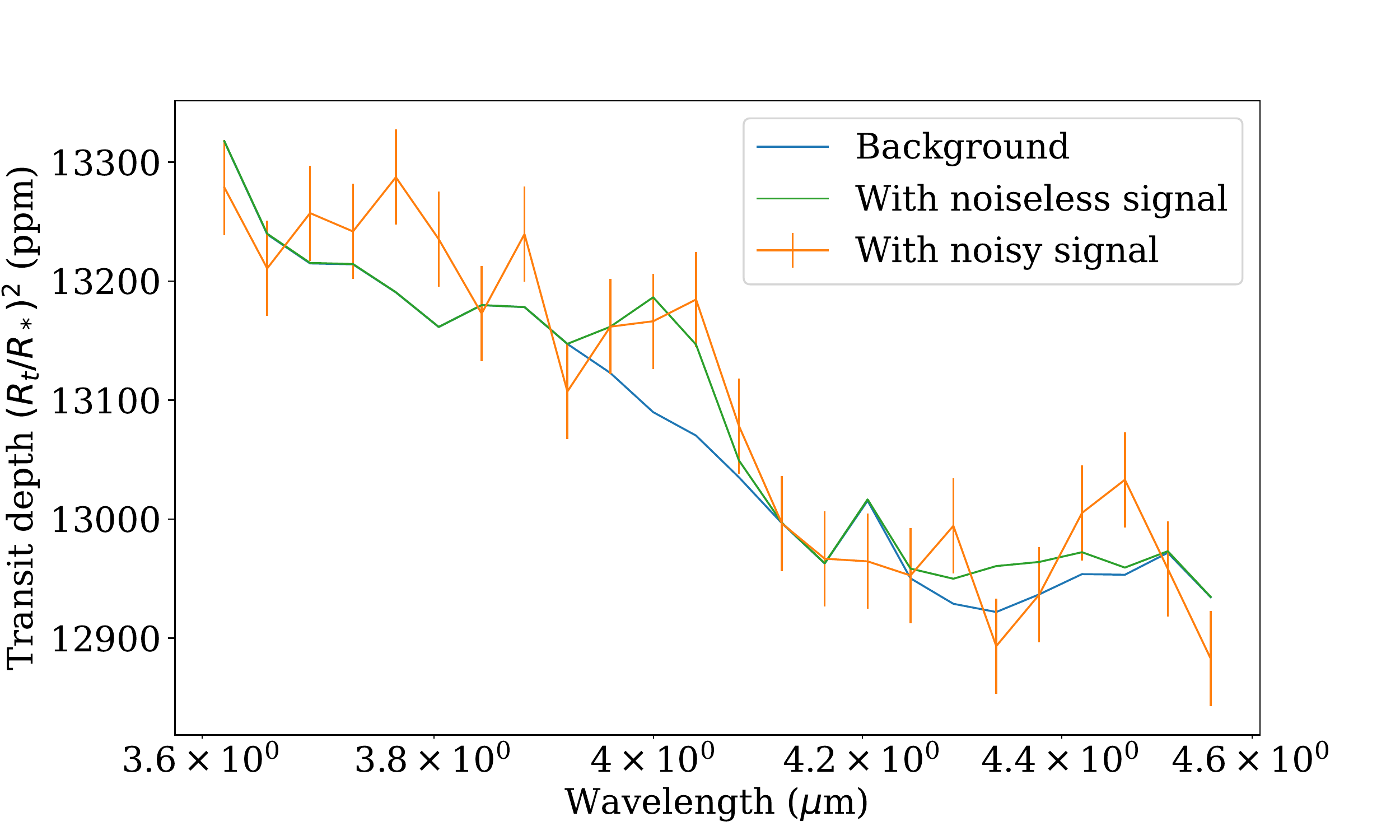}
    \includegraphics[width=0.45\textwidth]{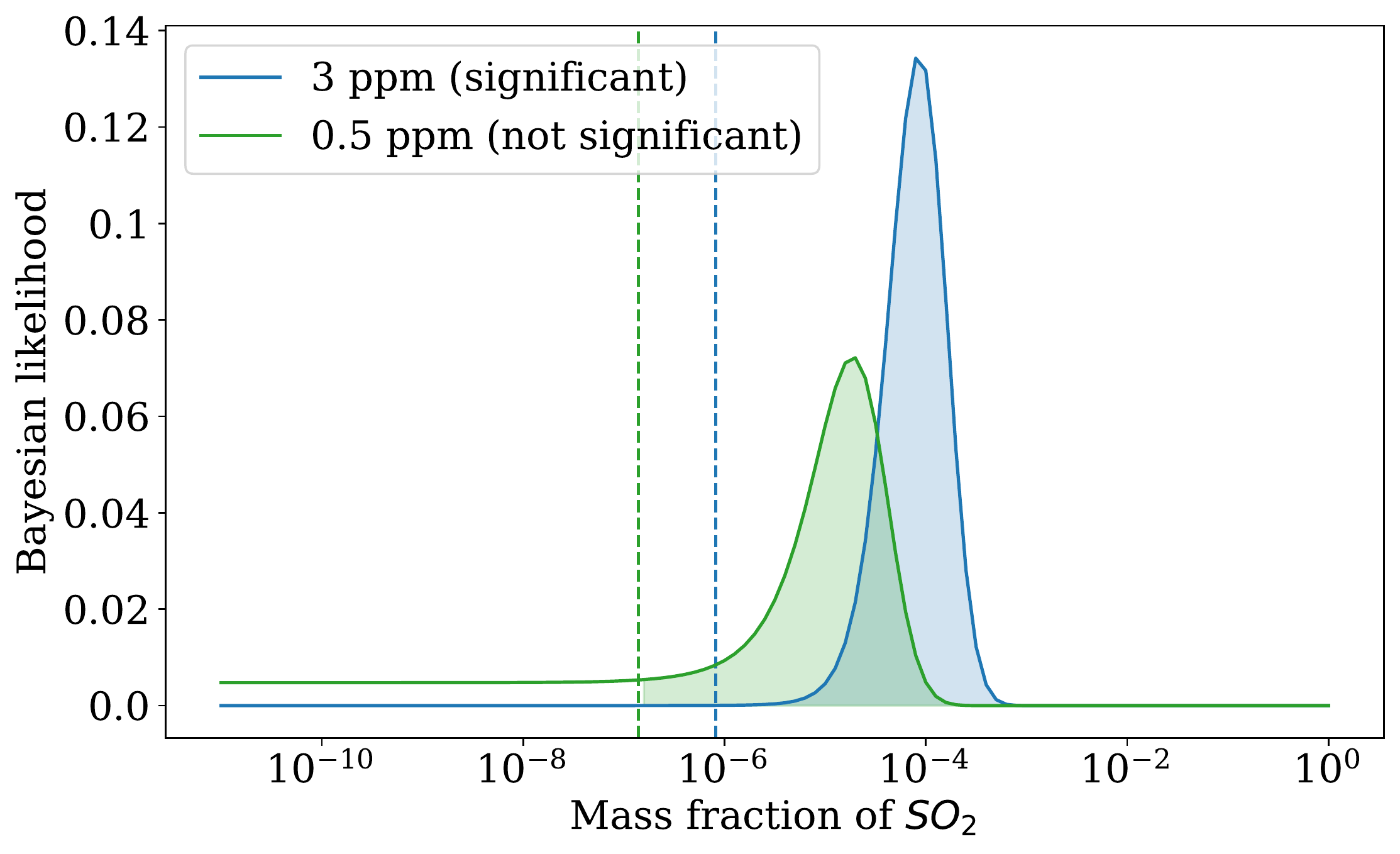}
    \caption{Simulated transmission spectra of a Hycean planet with added abundances of \ch{SO2} above the detection threshold (3 ppm, top left) and below the detection threshold (0.5 ppm, top right), and the \ch{SO2} mass fraction posterior PDF from detection tests (bottom). The spectral feature at 4 $\mu\mathrm{m}$ is targeted by the detection test. The spectra are presented with and without simulated \textit{JWST} noise, which is used for the detection test. The 1\% input, above which the posterior PDF is integrated to calculate the significance of detection, is highlighted with a dotted line in the detection test.}
    \label{fig:Retrieval}
\end{figure}

The mean and standard deviation of the retrieved abundance can then be calculated from the posterior PDF. The maximum abundance of the species present can be constrained by integrating the PDF: the lowest abundance that when integrated up to gives a value of 99.7\% (3$\sigma$) is the maximum abundance. To estimate the significance of a detection, the posterior PDF is integrated above a certain fraction of the input (we use 0.01).  This is equivalent to working out the probability that you retrieve 1\% or more of the abundance you input into the forward model. If this is equal to or greater than 99.7\% we label the detection as significant at the 3$\sigma$ level. The detection threshold is found by adding progressively more of a prebiosignature species until a significant detection is achieved. For an illustrative example of this see \hyperref[fig:Retrieval]{Figure 4}.

\sed{TriArc detection tests can utilise the entire wavelength range of an instrument to calculate the detection threshold of a prebiosignature, combining information from all spectral features. Alternatively, it can narrowly target individual spectral features, which are listed in \hyperref[tab:bands]{Table 3}.}

\sed{Using the entire wavelength range of the instrument gives the most optimistic detection thresholds, quoted in \hyperref[tab:planet thresholds]{Table 4}, and more closely matches the methodology of a full retrieval. Targeting individual features results in more conservative detection thresholds, and illustrates the relative strength of features in different backgrounds, and can be used to demonstrate how molecules which share their strongest feature can be distinguished. We do this by defining a disambiguation threshold, the minimum abundance required to distinguish between molecules with overlapping features, which is equal to the detection threshold of a molecule targeted at the strongest non-overlapping band. The results of this method are presented in \hyperref[fig:G395M]{Figure 5} and subsequent figures.}


\begin{deluxetable}{cccc}
\label{tab:bands}
\tablehead{\colhead{Wavelength} & \colhead{\sed{Prebiosignature}} & \colhead{\sed{Other}}  & \colhead{}\\ 
\colhead{Range ($\mathrm{\mu m}$)} & \colhead{Molecules} & \colhead{Molecules} & \colhead{Instrument}}
\tablecaption{Spectral Bands of Prebiosignature Molecules}
\startdata
1.55-1.65 & \ch{H2S}, \ch{NH3} & & \sed{NIRISS}\\
1.90-2.00 & \ch{H2S} & \ch{H2O}, \ch{CO2} & \sed{NIRISS}\\
2.10-2.30 & \ch{NH3}, \ch{CH4} &  &  \sed{NIRISS}\\
2.65-2.75 & \ch{NO}, \ch{H2S} & \ch{H2O}, \ch{CO2} & \sed{NIRISS}\\
2.90-3.10 & \ch{HCN}, \ch{C2H2}, \ch{NH3}, \ch{HC3N} & \ch{CO2} &  G395M\\
3.30-3.50 & \ch{CH4}, \ch{CH2O} &  &  G395M\\
3.50-3.60 & \ch{HCN}, \ch{CH4} &   &  G395M\\
3.65-3.80 & \ch{C2H2}, \ch{H2S} & &  G395M\\
3.90-4.00 & \ch{SO2}, HCN & &  G395M\\
4.20-4.50 & \ch{HC3N}, \ch{SO2} & \ch{CO2} &  G395M\\
4.55-4.65 & \ch{C2H2} &  &  G395M\\
4.70-4.80 & \ch{HCN}, CO &  &  G395M\\
4.80-5.00 & \ch{HC3N}, CO & &  G395M\\
5.00-6.00 & \ch{NO} & \ch{H2O} &  MIRI\\
6.20-6.90 & \ch{NH3} & \ch{H2O} &  MIRI\\
7.10-7.80 & \ch{HC3N}, \ch{SO2}, \ch{C2H2}, \ch{HCN}, \ch{H2S} & & MIRI\\
\enddata

\end{deluxetable}

\section{Results}
\label{sec:Results}

\begin{figure}
    \centering
    \includegraphics[width=0.65\textwidth]{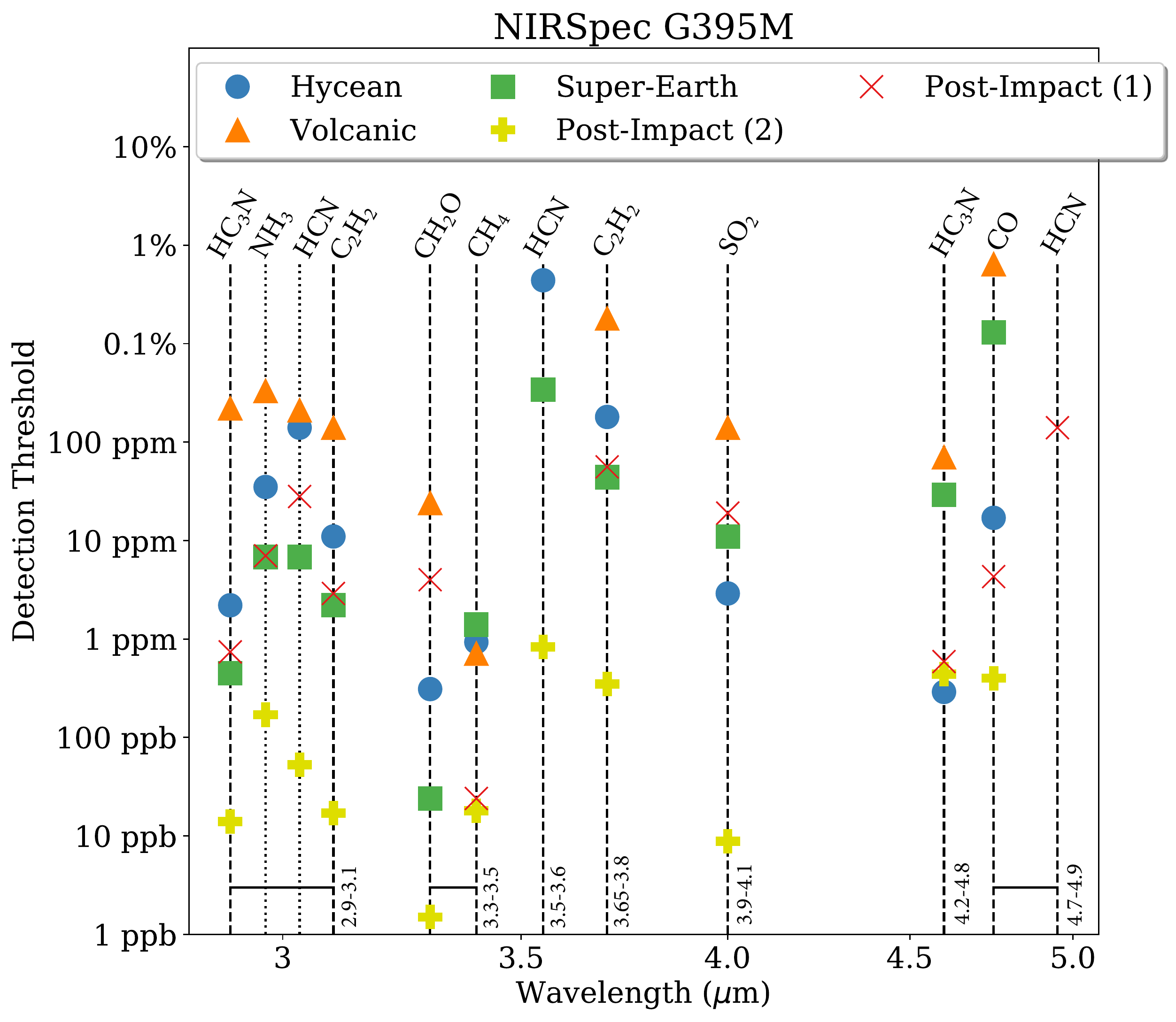}
    \includegraphics[width=0.3455\textwidth]{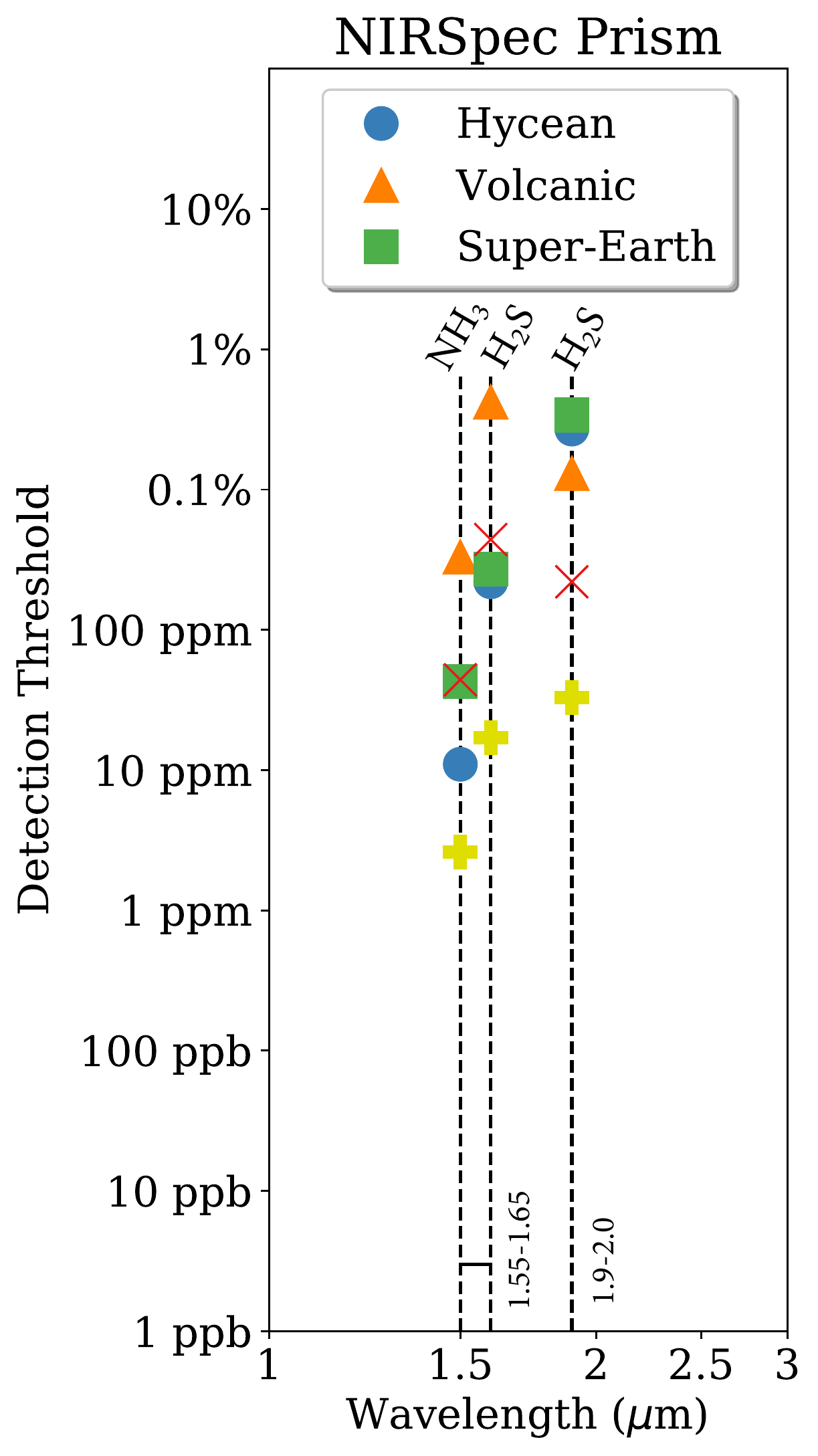}
    \includegraphics[width=0.4545\textwidth]{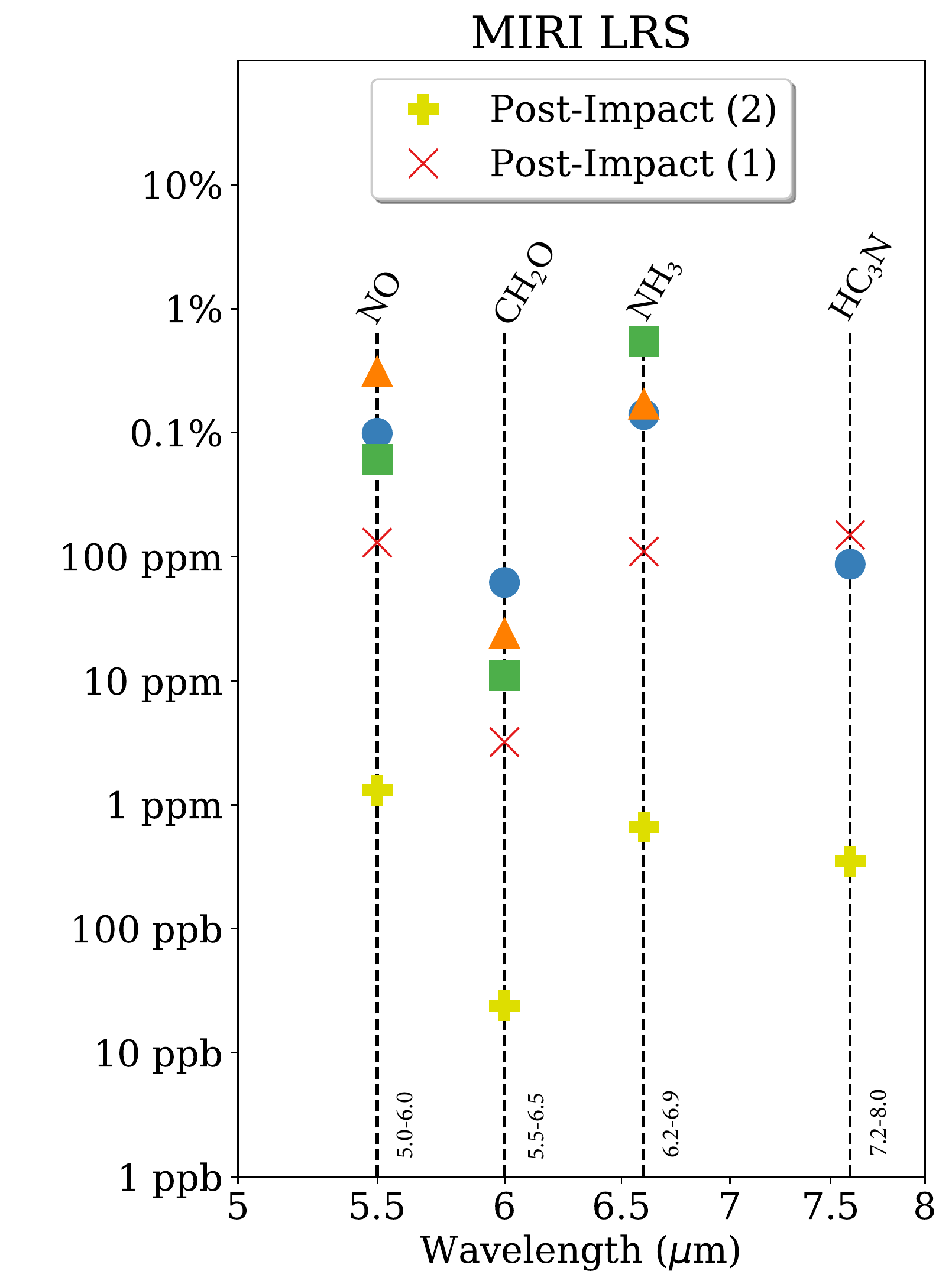}
    \caption{Detection thresholds of prebiosignature molecules at different wavelengths for different model exoplanets. \textit{JWST} noise is simulated using three transits per instrument of GJ 1132b at spectral resolution R=100. Post-Impact (1) refers to the 100 kyr Post-Impact case and Post-Impact (2) refers to the 10 Myr Post-Impact case.}
    \label{fig:G395M}
\end{figure}

Here we present a selection of prebiosignature detection thresholds summarised in \hyperref[fig:G395M]{Figure 5}. For each model exoplanet we calculate the detection threshold using a consistent synthetic noise profile based on observing three transits of GJ 1132b (see \hyperref[subsec:PandExo]{Section 2.2}) to demonstrate the dependence of detection thresholds on planetary parameters. These results are listed in \hyperref[subsec:Thresholds]{Section 3.1}. \sed{We include additional results with added grey cloud decks at various altitudes to quantify the impact of clouds in \hyperref[subsec:clouds]{Section 3.2}. We benchmark these detection thresholds results against full retrieval results with {\tt petitRADTRANS} in \hyperref[subsec:trueretrieval]{Section 3.3}.} We also explore the impact of noise profiles from alternative observational regimes on the detection thresholds in \hyperref[subsec:observational]{Section 3.4}. To quantify the observational regime necessary for prebiosignature analysis, we calculate the detection thresholds as a function of number of observed transits using a model Hycean planet as a benchmark. We extend this by testing the impact of observing different stars for a varying number of transits on the HCN detection threshold in the Hycean benchmark planet. We also do this for our model TRAPPIST-1e with high mean molecular weight (90\% \ch{N2}, 10\% \ch{CO2}) atmosphere. In \hyperref[subsec:degeneracies]{Section 3.5} we discuss the handling of degeneracies between different prebiosignatures; we introduce disambiguation thresholds, the minimum abundances to distinguish between molecules. The detection thresholds are compiled in \hyperref[tab:planet thresholds]{Table 4} and illustrated in \hyperref[fig:G395M]{Figure 5}. We explore the sensitivity of our results to temperature, mean molecular weights, planet radius, surface pressure, grey cloud layers, star radius, and instrument precision in \hyperref[sec:sensitivity]{Appendix A}.

\subsection{Detection Thresholds}
\label{subsec:Thresholds}

\begin{deluxetable}{ccccccc}[ht!]
\label{tab:planet thresholds}
\tabletypesize{}
\tablehead{\colhead{Molecule} & \colhead{Instrument} & \colhead{Hycean} & \colhead{Ultrareduced Volcanic} & \colhead{Super-Earth} & \colhead{Post-Impact (100 kyr)} & \colhead{Post-Impact (10 Myr)}\\
\colhead{} &  & \colhead{(ppm)} & \colhead{(ppm)} & \colhead{(ppm)} & \colhead{(ppm)} & \colhead{(ppm)}}
\tablecaption{Detection Thresholds of Prebiosignature Molecules in Model Exoplanet Atmospheres, Using Three Transits Per Instrument of GJ 1132b at Spectral Resolution R=100}

\startdata
\ch{CH4} & G395M & 0.37 & 0.28 & 0.46 & 0.0094 & 0.0089\\
HCN & G395M & 27 & 66 & 1.7 & 4.4 & 0.026\\
\ch{C2H2} & G395M & 3.6 & 55 & 0.56 & 1.4 & 0.0087\\ 
\ch{NH3} & G395M & 5.5 & 84 & 2.2 & 2.8 & 0.084\\
& NIRISS & 0.55 & 3.3 & 0.68 & 0.88  & 0.042 \\
& MIRI &  69   & 53 & 14 & 8.8  & 0.084\\
\ch{SO2} & G395M & 0.58 & 14 & 0.45 & 3.7 & 0.028\\
\ch{H2S} & G395M & 110 & 2100 & 27 & 280 &  1.3\\
& NIRISS & 69 & 130 & 54 & 35 & 4.2\\
NO & MIRI & 360 & 870 & 61 & 40 & 0.40\\ 
CO & G395M & 1.3 & 100 & 2.1 & 0.54 & 0.051 \\
\ch{CH2O} & G395M & 0.16 & 9.5 & 0.024 & 2.0  & 0.00075\\
& MIRI & 16 & 7.5  & 3.1  & 1.3 & 0.012\\
\ch{HC3N} & G395M & 0.058 & 5.6 & 0.14 & 0.15 & 0.0070\\
& MIRI & 87 & - & - & 150  & 0.35\\
\ch{NH3} & G395M & 35  & 330 & 6.8 & 7.0 & 0.17\\
\enddata

\end{deluxetable}

For our primary observational regime, we calculate the detection thresholds by simulating the transmission spectrum of three transits around GJ 1132 at a spectral resolution of R=100, as described in \hyperref[subsec:PandExo]{Section 2.2}, for each of the following instruments: \sed{NIRISS SOSS}, NIRSpec G395M, and MIRI LRS (for a total of six hours of observation per instrument). We find this to be a good compromise between observation time and detection thresholds. This is done for all of the model hydrogen-rich exoplanets. A summary of these results is found in \hyperref[tab:planet thresholds]{Table 4} and \hyperref[fig:G395M]{Figure 5}. No prebiosignature molecules in the model early-Earth exoplanet are detectable with this regime.


Using the primary observation regime, all of the prebiosignatures are detectable in the model Hycean world, with its \ch{CH4}- and \ch{H2O}-dominated transmission spectrum. Abundances of 0.06 ppm \ch{HC3N}, 0.6 ppm \ch{SO2}, 4 ppm \ch{C2H2}, 1.3 ppm CO and 30 ppm HCN are detectable using the NIRSpec G395M instrument. 70 ppm \ch{H2S} is detectable by the \sed{NIRISS SOSS} instrument, and a mixing ratio of at least 400 ppm \ch{NO} is detectable using MIRI LRS. \ch{CH4} and \ch{NH3} are \sed{present in chemical equilibrium in this atmosphere, and are hence considered as background species}. The spectral feature of \ch{CH2O} at 6 $\mathrm{\mu m}$, required to distinguish it from \ch{CH4}, is detectable at 16 ppm with MIRI LRS.  We also use the Hycean planet as a benchmark for alternative observational regimes (\hyperref[subsec:observational]{Section 3.2}).


The hydrogen-rich super-Earth, with its \ch{CO2} and \ch{H2-H2} transmission spectrum, is very well-suited to most prebiosignatures molecules, despite it having the lowest scale height of the planets we consider, due to the lack of absorbing molecules in the important region around 3 $\mu\mathrm{m}$. Abundances of 0.02 ppm of \ch{CH2O}, 0.14 ppm of \ch{HC3N}, 0.5 ppm of \ch{CH4}, 0.6 ppm \ch{C2H2}, 7 ppm \ch{NH3}, 2 ppm of HCN, 0.5 ppm \ch{SO2}, and 2 ppm of CO, and 30 ppm of \ch{H2S} are all detectable with NIRSpec G395M. \ch{NO} is detectable at 60 ppm with MIRI LRS. Our resultants are concordant with \citet{huang2021assessment}, which finds a detection threshold of 5 ppm for ammonia in a hydrogen-dominated super-Earth atmosphere.


The ultrareduced volcanic planet, despite having a similar scale height to the Hycean world, has generally worse detection thresholds due to high concentrations of strongly-absorbing \ch{CH4} and \ch{HCN} in the atmosphere. In a clear atmosphere, abundances of 6 ppm \ch{HC3N}, 14 ppm of \ch{SO2},  60 ppm of \ch{C2H2}, and 80 ppm \ch{NH3} are detectable with NIRSpec G395M. \ch{CH4} and \ch{HCN} are part of the background atmosphere due to the ultra-reduced outgassing. \ch{NO} is detected with MIRI LRS for an abundance of at least 900 ppm, and the 6 $\mathrm{\mu m}$ feature of \ch{CH2O} for an abundance of 8 ppm. A 130 ppm mixing ratio of \ch{H2S} is detectable with \sed{NIRISS SOSS}, and a 100 ppm mixing ratio of CO with NIRSpec G395M.


The post-impact planets, with their high temperature, hydrogen-dominated atmospheres, have the highest scale height, and are hence the best-suited to detection. In the \ch{CH4}-dominated spectrum of the 100 kyr post-impact Hadean Earth, we could detect 150 ppb \ch{HC3N}, 1.4 ppm \ch{C2H2}, 0.5 ppm CO, 7 ppm \ch{NH3}, 4 ppm \ch{SO2} and 4 ppm \ch{HCN} using NIRSpec G395M, 40 ppm \ch{H2S} using \sed{NIRISS SOSS}, and 1.3 ppm \ch{CH2O} and 40 ppm NO using MIRI LRS.

The 10 Myr post-impact transmission spectrum is instead dominated by the lower opacities of \ch{CO} and \ch{H2-H2}, resulting in low detection thresholds that enable trace abundances of prebiosignatures to be detected: 0.8 ppb \ch{CH2O}, 7 ppb \ch{HC3N}, 9 ppb \ch{CH4} and \ch{C2H2}, 26 ppb \ch{HCN}, 30 ppb \ch{SO2}, and 1.3 ppm of \ch{H2S} all using NIRSpec G395M. Also detectable is 0.4 ppm of \ch{NO} with MIRI LRS.

\subsection{Impact of Clouds}
\label{subsec:clouds}
\sed{In the case of both the hydrogen-rich super-Earth and the 10 Myr post-impact planet, the overall opacity is very low, with the transmission spectrum reaching deep in the planetary atmospheres to where the continuum \ch{H2-H2} opacity becomes optically thick. In this case, any source of higher-altitude opacity, including clouds and aerosols, would massively impact the transmission spectrum and therefore the detection thresholds. The other spectra, which already possess strongly absorbing broadband opacity sources from \ch{CH4}, \ch{NH3}, \ch{H2O}, and \ch{HCN}, would be less impacted by clouds, but it would still affect detection thresholds.}

\sed{To quantify the impact of various cloud heights on these detection thresholds, we rerun TriArc with grey cloud decks at various altitudes in various planets. Species with their strongest features in unsaturated regions (like CO in \ch{CH4}-dominated atmospheres) are more steeply affected than species in saturated regions (like \ch{C2H2} in atmospheres containing \ch{NH3}). For the hydrogen-rich super-Earth we add a cloud deck at 0.1 bar to represent Earth-like water clouds, and find that detection thresholds increase by a factor of 10--20. For the ultra-reduced volcanic planet, using the same aerosol layer at 10 mbar as \citet{rimmer2021detectable} causes a 3--10 times increases in the detection thresholds. To represent a high-altitude organic haze in the 100 kyr post-impact atmosphere we use a 1 mbar cloud deck, greatly flattening the background transmission spectrum and increasing most detection thresholds between 3--20 times (3 for \ch{C2H2}, 4 for \ch{SO2}, 10 for HCN, 20 for NO), yet 1000 times for CO. For our most extremely cloudy scenario, we use an extremely high-altitude aerosol at 100 $\mu$bar cloud deck in the 10 Myr post-impact atmosphere. This creates an effectively flat background transmission spectrum, and the corresponding detection thresholds are purely an expression for the strength of the molecule's strongest feature. Therefore detection thresholds of strong absorbers like \ch{HC3N}, \ch{C2H2}, and \ch{CH4} only increase by 100--200 times, while weaker absorbers like \ch{SO2}, \ch{H2S}, and HCN increase by 2500--6000 times, and both CO and NO become undetectable.}

\subsection{Retrieval Results}
\label{subsec:trueretrieval}

As a consequence of our detection test method, the detection thresholds may not necessarily correspond to the exact minimum abundances that we might obtain from real observations. We may expect our results to be optimistic as we assume perfect knowledge of the exoplanet and the other species in its atmosphere. In a realistic retrieval, any uncertainty in the atmospheric composition or planetary properties will affect the retrieved abundance of trace species. As we are considering detection of trace prebiosignatures, it \sed{is reasonable} to assume that in practice such observations will be made for planets where the dominant absorbing gases in their atmospheres have already been well constrained.

\begin{figure}[ht]
    \centering
    \includegraphics[width=0.97\textwidth]{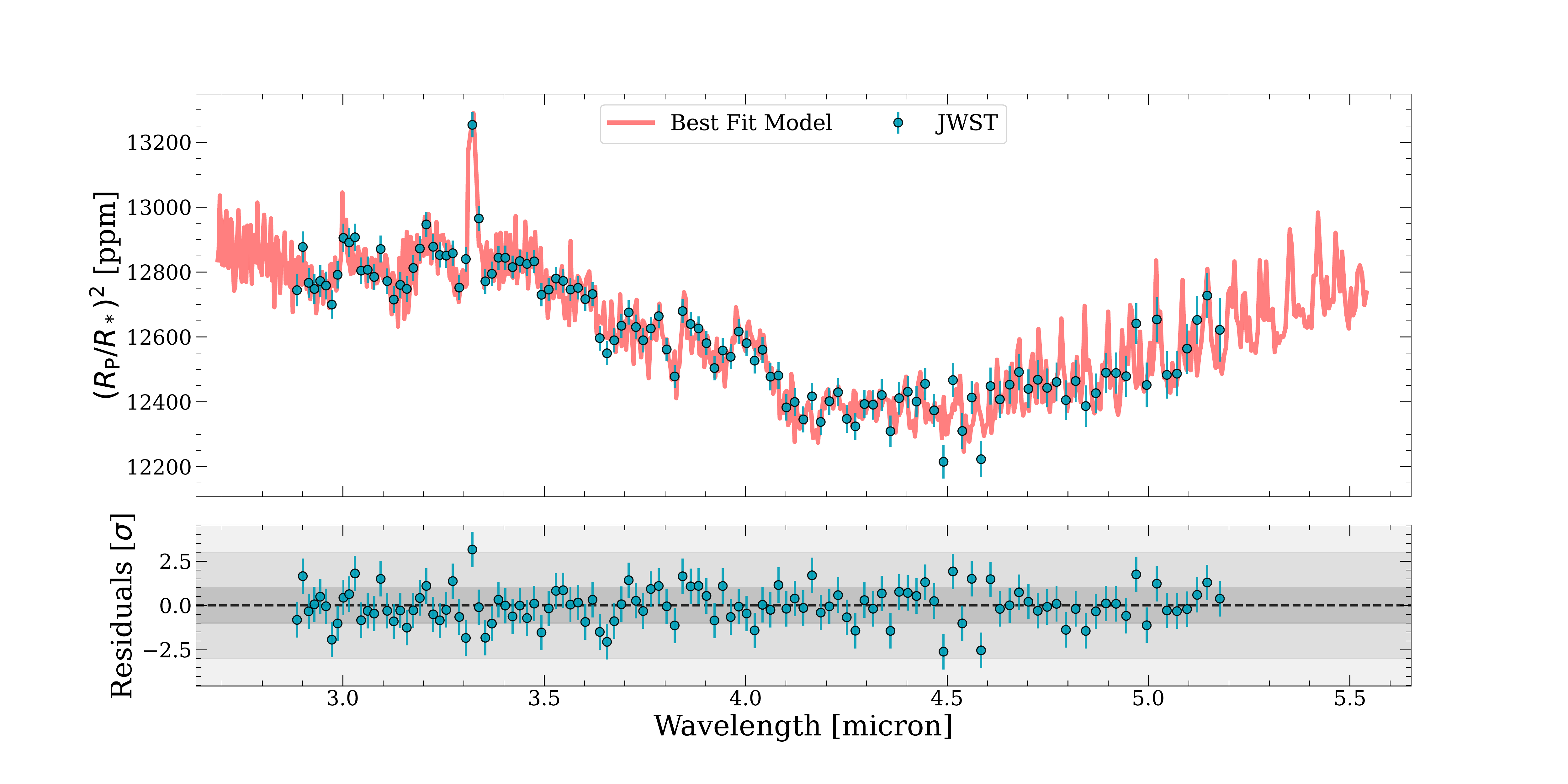}
    \caption{\sed{Best fit transmission spectrum to a simulated NIRSpec G395M observation of a Hycean atmosphere with {\tt TriArc}-calculated detection thresholds abundances of three prebiosignatures (HCN, \ch{C2H2}, and \ch{SO2}) assuming realistic JWST noise from observing three transits of GJ 1132b, calculated using the retrieval package of {\tt petitRADTRANS}.}}
    \label{fig:best_fit_spectrum}
\end{figure}

\begin{figure}[p]
    \centering
    \includegraphics[width=1.0\textwidth]{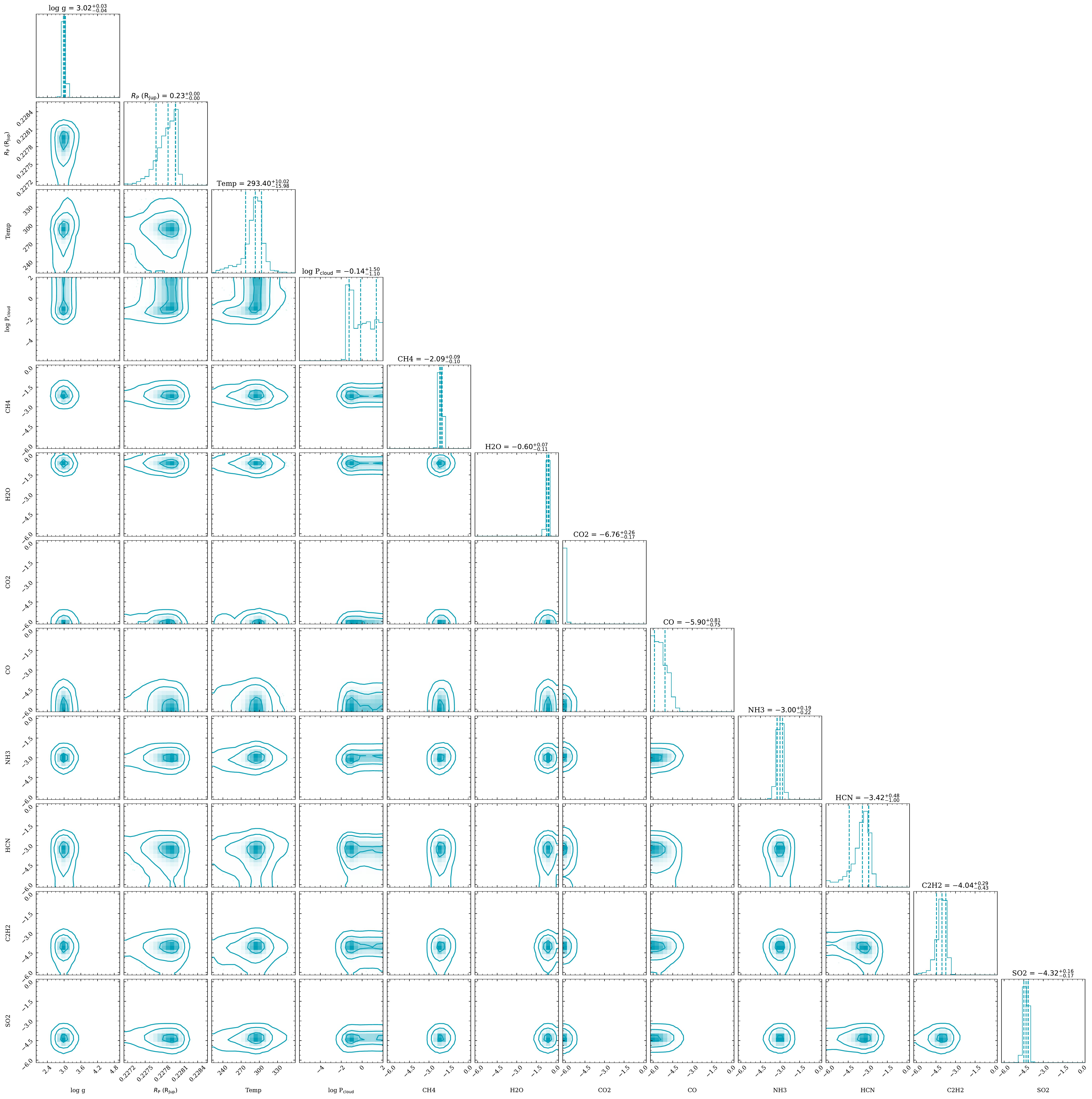}
    \caption{\sed{Full posterior distribution from the {\tt petitRADTRANS} retrieval of a simulated NIRSpec G395M observation of a Hycean atmosphere with {\tt TriArc}-calculated detection thresholds abundances of three prebiosignatures (HCN, \ch{C2H2}, and \ch{SO2}) assuming realistic JWST noise from observing three transits of GJ 1132b. The prebiosignature mass fractions are identified within 2$\sigma$ of their true values (HCN = -2.78, \ch{C2H2} = -3.92, \ch{SO2} = -4.1). Due to the overlapping strongest features of HCN, \ch{C2H2}, and \ch{NH3}, there is a tail in the posterior PDF of both HCN and \ch{C2H2} as the interpretation for the shared feature at 3 $\mathrm{\mu m}$ is somewhat degenerate.}}
    \label{fig:trueretrieval}
\end{figure}

\sed{To verify that prebiosignatures of detection threshold abundances can still be identified in a realistic retrieval, we perform a full retrieval using the retrieval package of {\tt petitRADTRANS} on a simulated NIRSpec G395M observation of a Hycean atmosphere. We use the planetary properties and observational regime described in \hyperref[sec:Method]{Section 2}, with added abundances of the prebiosignatures \ch{HCN}, \ch{C2H2}, and \ch{SO2} equal to their detection thresholds (27 ppm, 3.6 ppm, and 0.58 ppm respectively). The results of the retrieval (see \hyperref[fig:best_fit_spectrum]{Figure 6} for the best fit spectrum and \hyperref[fig:trueretrieval]{Figure 7} for the posterior corner plot) successfully identify all three prebiosignatures at their correct abundances (within 2$\sigma$). The retrieval accurately constrains the planetary parameters and background composition (\ch{H2O}, \ch{CH4}, and \ch{NH3}), and places upper limits on the \ch{CO} and \ch{CO2} abundances, which aren't present in the forward model. The retrieval highlights the degeneracy present between \ch{C2H2} and \ch{HCN} evident as a weakly correlated tail in the posterior PDF, as the interpretation for the feature at 3$\mathrm{\mu m}$ is degenerate and can be explained with either molecule. The \ch{SO2} feature is non-degenerate, and its posterior PDF is therefore better constrained.}

\begin{figure}[p]
    \centering
    \includegraphics[width=1.0\textwidth]{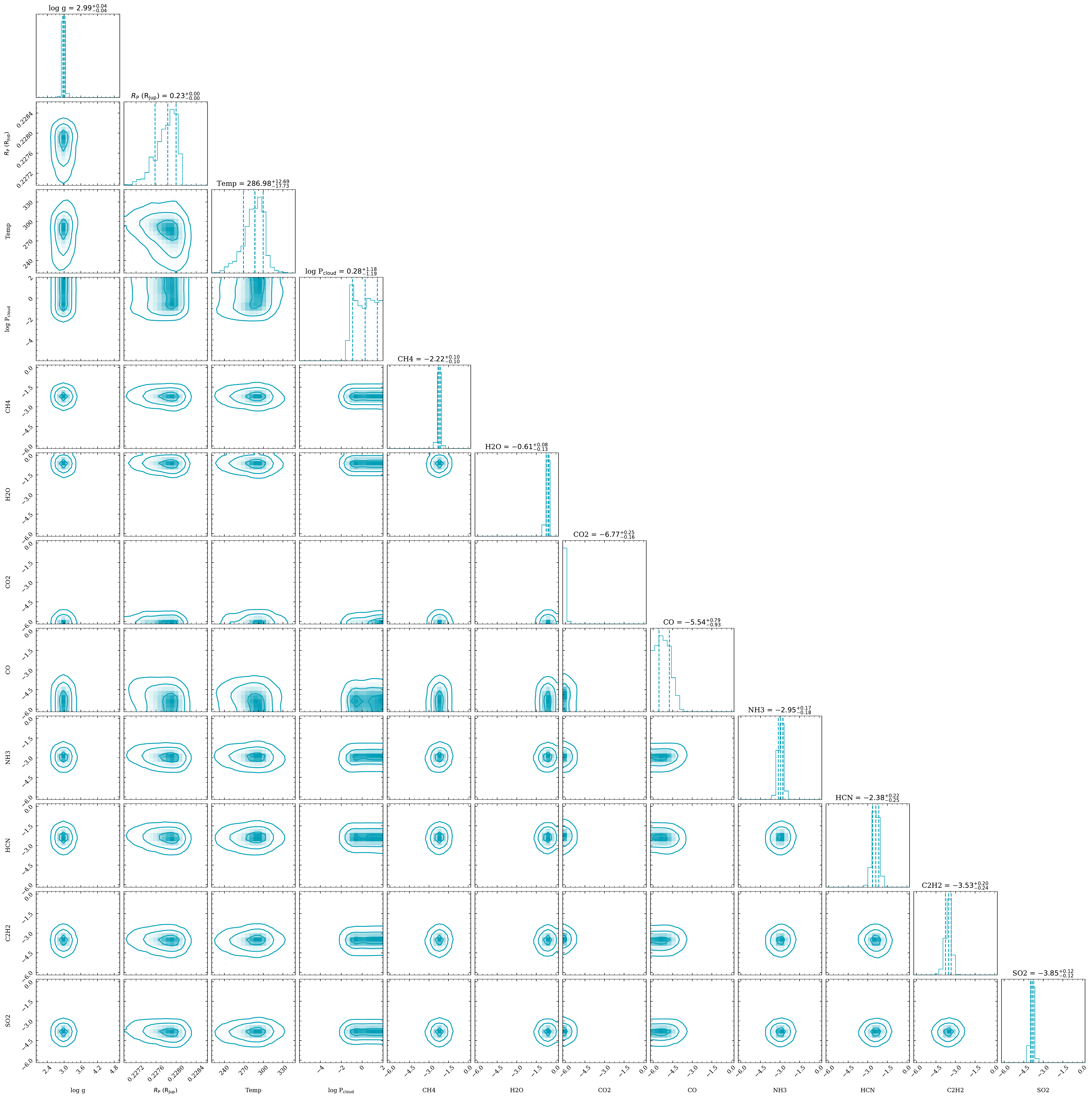}
    \caption{\sed{Full posterior distribution from the {\tt petitRADTRANS} retrieval of a simulated NIRSpec G395M observation of a Hycean atmosphere with 5 times the {\tt TriArc}-calculated detection thresholds abundances of three prebiosignatures (HCN, \ch{C2H2}, and \ch{SO2}) assuming realistic JWST noise from observing three transits of GJ 1132b. The prebiosignature mass fractions are identified within 1--2$\sigma$ of their true values (HCN = -2.28, \ch{C2H2} = -3.42, \ch{SO2} = -3.6). With the increased abundances all three prebiosignatures are robustly detected and distinguished, demonstrating the order-of-magnitude accuracy of the detection threshold method.}}
    \label{fig:trueretrieval2}
\end{figure}

\sed{To demonstrate that the prebiosignatures can be robustly detected and distinguished within an order of magnitude of the TriArc-determined detection threshold, we repeat the retrieval, using 5 times the abundance of the prebiosignatures (HCN, \ch{C2H2}, and \ch{SO2}). The posterior corner plot from this retrieval is presented in \hyperref[fig:trueretrieval2]{Figure 8}. With the increased prebiosignature abundances, all three prebiosignatures are detected and distinguished with tight abundance constraints. Overlapping bands are therefore demonstrated to be a problem at the detection threshold, but the appearance of non-shared spectral features at slightly higher abundances implies that the disambiguation threshold, even for the highly similar molecules HCN and \ch{C2H2}, is within an order-of-magntitude of the detection threshold.}


\subsection{Alternative Observational Regimes}
\label{subsec:observational}

\begin{figure}[p]
    \centering
    \includegraphics[width=0.8\textwidth]{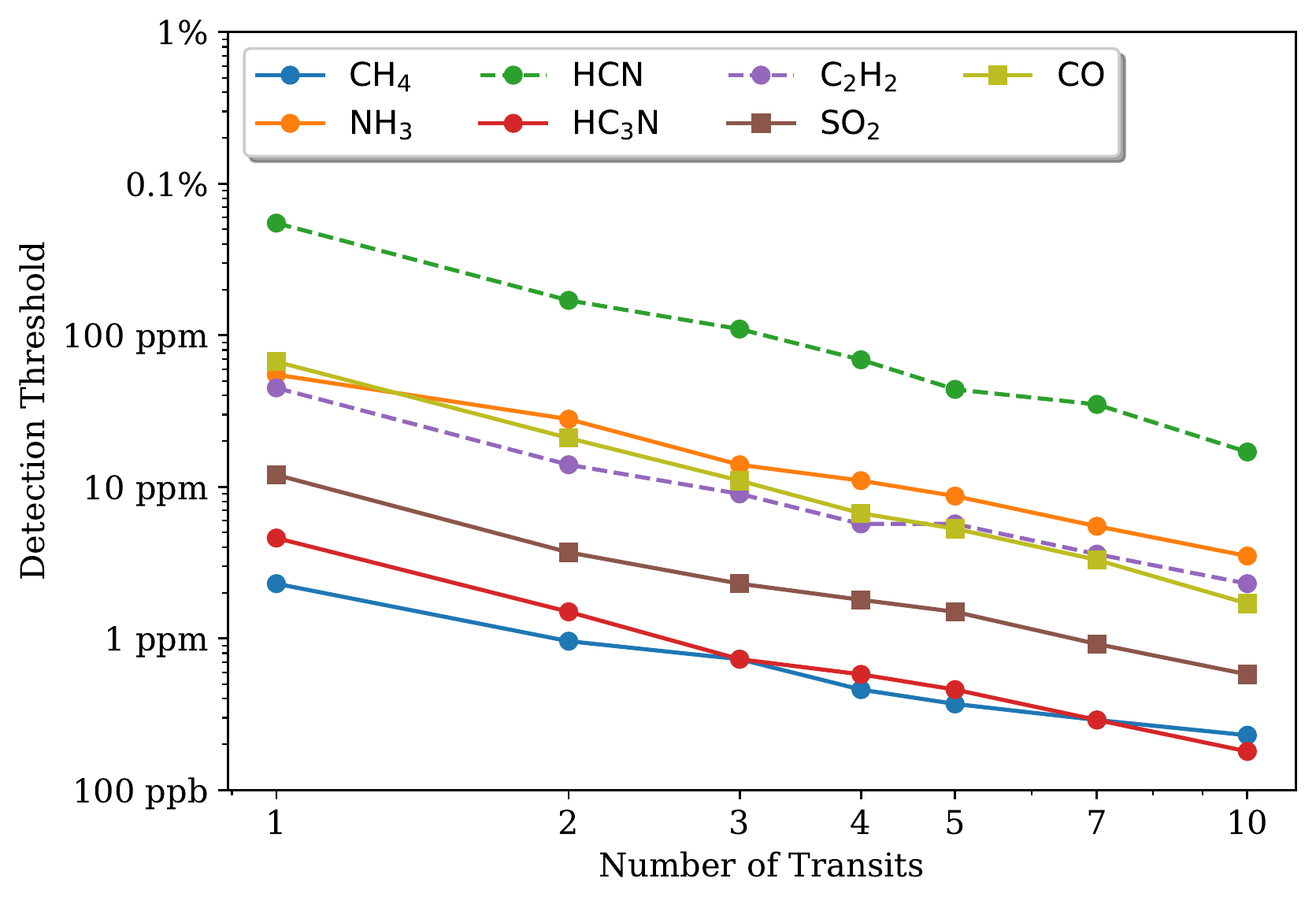}
    \caption{Detection thresholds for prebiosignature molecules with NIRSpec G395M as a function of number of transits using simulated transmission spectra of a model Hycean exoplanet, with synthetic JWST noise from observing GJ 1132 at spectral resolution R=100.}
    \label{fig:transitshycean}
\end{figure}

\begin{figure}[p]
    \centering
    \includegraphics[width=0.8\textwidth]{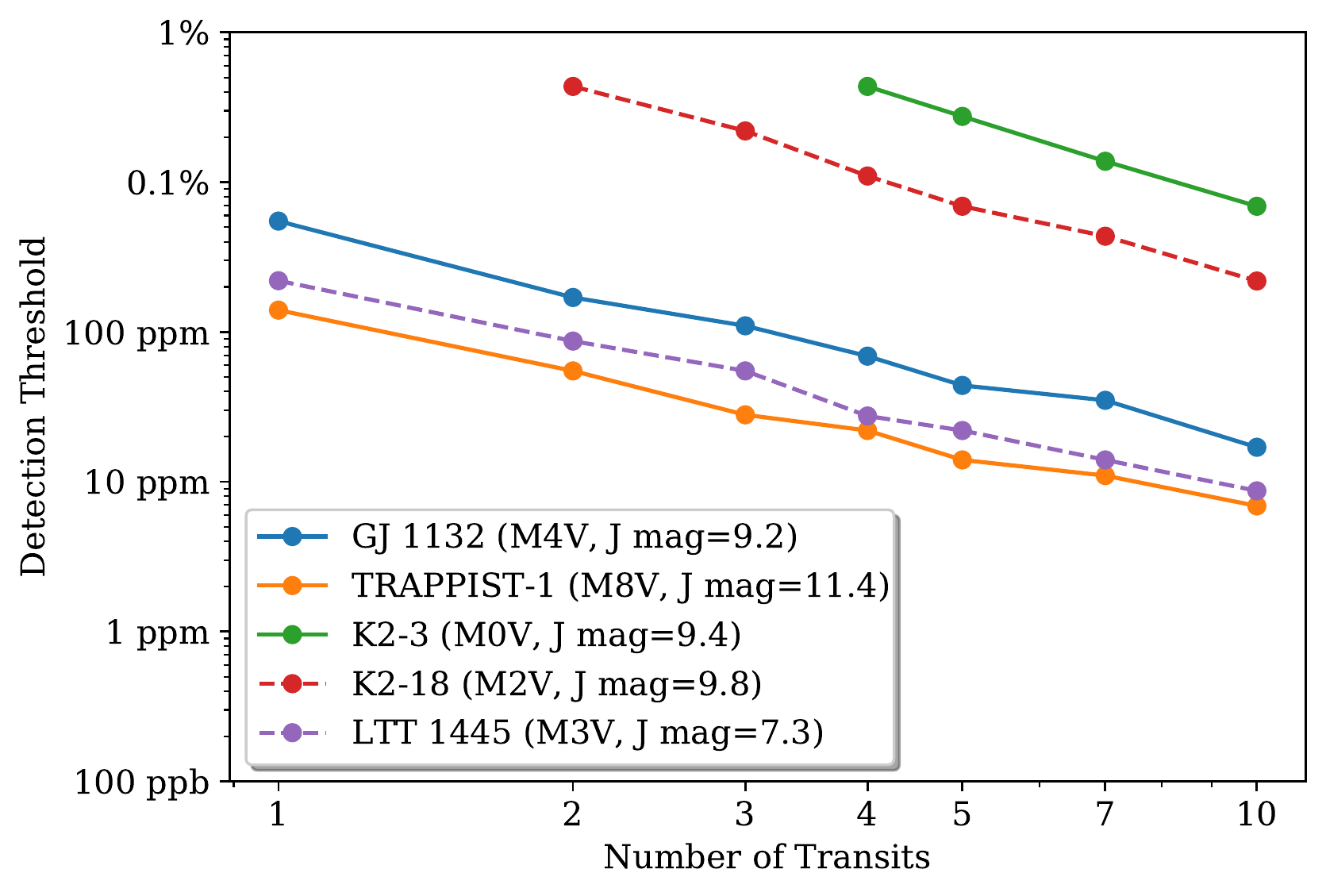}
    \caption{Detection thresholds of HCN as a function of number of transits for a selection of model stars, using simulated transmission spectra of a model Hycean exoplanet at spectral resolution R=100.}
    \label{fig:transitsstars}
\end{figure}

We here seek to generalize the analysis of detection thresholds by varying some of the assumed observational parameters. To begin with, we calculate the detection thresholds while varying the number of transits, but otherwise keeping the other observational parameters the same as in the primary regime (GJ 1132 with a three hour baseline). This is done for the Hycean benchmark planet with the detection thresholds of \ch{CH4}, \ch{NH3}, HCN, \ch{HC3N}, \ch{C2H2}, \ch{SO2}, and CO, all using the NIRSpec G395M instrument (\hyperref[fig:transitshycean]{Figure 9}). All of these prebiosignatures are detectable in a single transit. Detection thresholds reduce by a factor of 10-40 by going from one to ten transits.

We also want to explore the impact of observing different stars. As the strength of the signal in transmission spectroscopy is proportional to $\frac{1}{R_*^2}$, only the smallest stars are suitable to atmospheric characterization with \textit{JWST}. Furthermore, observing brighter stars results in less noise. To quantify these effects, we use noise from a sample of relevant planetary systems: GJ 1132b, TRAPPIST-1e, LTT 1445 Ab, K2-3d, and K2-18b, to calculate the detection threshold of HCN in the model Hycean planet with a variable number of transits (\hyperref[fig:transitsstars]{Figure 10}). In each case, we use a baseline observing time equal to the duration of three transits for that particular planetary system. As expected, the detection threshold strongly depends on stellar radius, and also depends on the star's magnitude at relevant wavelengths. Regardless of the type of star the detection threshold decreases by approximately the same relative amount with number of transits. HCN is detectable in Hycean LTT 1445 Ab, GJ 1132b, and TRAPPIST-1e with a single transit, two transits are required for K2-18b, and four transits for K2-3d.

\begin{figure}
    \centering
    \includegraphics[width=0.6\textwidth]{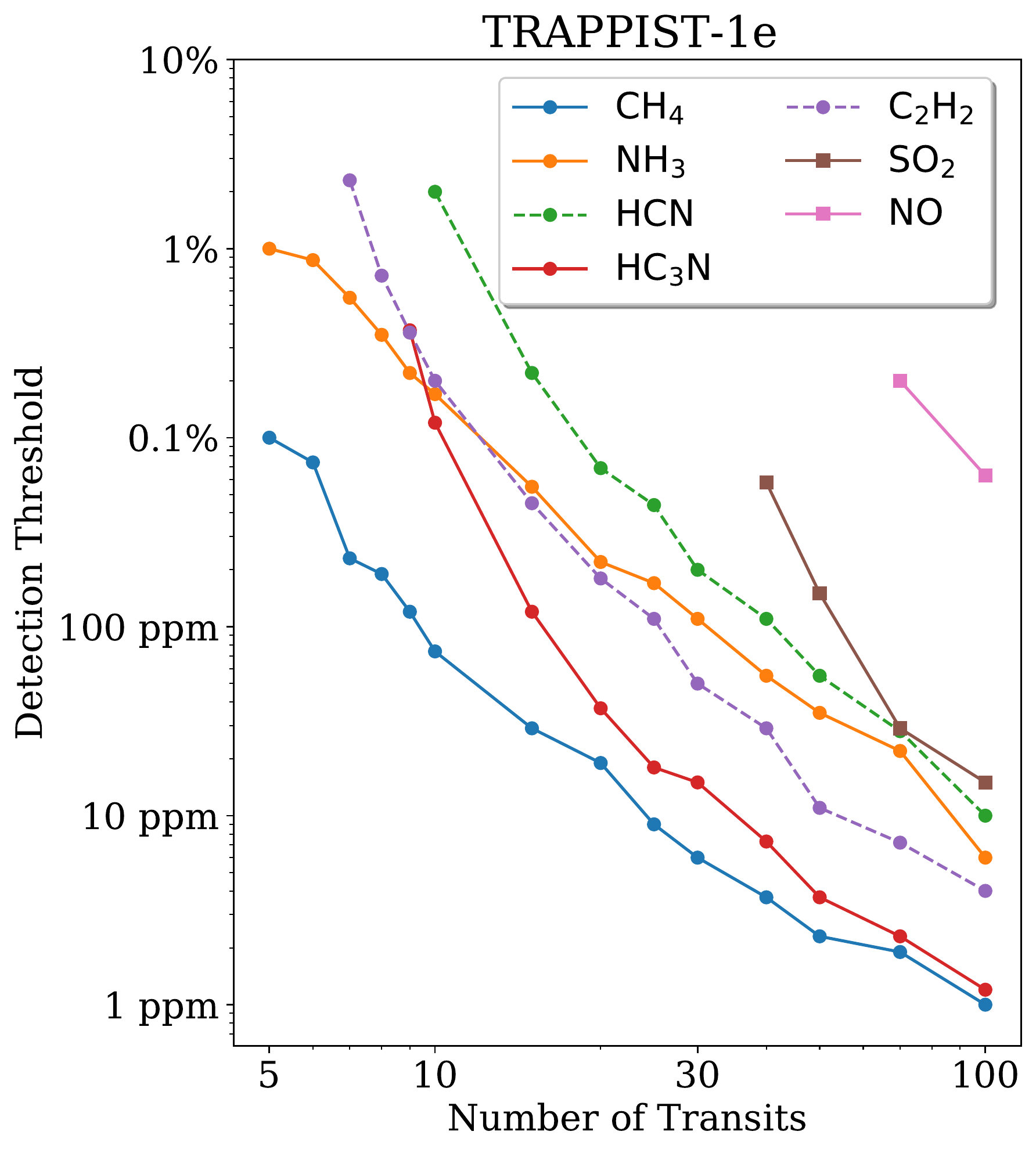}
    \caption{Detection thresholds of prebiosignature molecules as a function of number of transits using simulated transmission spectra of TRAPPIST-1e with an early Earth atmosphere at spectral resolution R=100.}
    \label{fig:thresholdstrappist}
\end{figure}

We also want to test the efficacy of long duration observing regimes at detecting prebiosignatures in high mean molecular weight atmospheres. The habitable zone planet TRAPPIST-1e is ideally suited to observation and has been observed to lack a cloud-free low mean molecular weight atmosphere \citep{de2018atmospheric}. In order to explore both the high mean molecular weight atmosphere of early Earth and the TRAPPIST-1 system, we simulate surveys of 5 to 100 transits around M8V star TRAPPIST-1 at spectral resolution R=100 (each with 10 hours of out of transit observation). The noise we obtain from this is used to calculate detection thresholds with a varying number of transits for our model TRAPPIST-1e, with a high mean molecular weight 90\% \ch{N2} 10\% \ch{CO2} atmosphere to simulate the atmosphere of the Early Earth \citep{kaltenegger2007spectral,rugheimer2015uv}. We use a 10 ppm systematic noise floor for our high intensity regimes, which is approached but not reached at any point, even for 100 transits. These results are illustrated in \hyperref[fig:thresholdstrappist]{Figure 11}.

With 5 transits, we are able to detect only \ch{CH4} and \ch{NH3} both with the NIRSpec G395M instrument. This concurs with the analysis of \citet{morley2017observing}, who find the dominant absorbing gas is observable at four transits for TRAPPIST-1e. At these concentrations, both \ch{CH4} and \ch{NH3} would constitute the dominant absorbing gas. At seven transits we detect \ch{C2H2}, nine transits we detect \ch{HC3N}, and ten transits we detect HCN. We detect no more prebiosignatures until \ch{SO2} at 40 transits, \ch{NO} at 70 transits, and \ch{H2S} and CO at 100 transits.

\subsection{Degeneracies}
\label{subsec:degeneracies}

\ch{HCN} and \ch{C2H2} share the same strongest absorbing spectral feature, from 2.9--3.1 $\mathrm{\mu m}$. This results in a degeneracy, as the presence of \ch{HCN} would result in the detection of \ch{C2H2} and vice versa using {\tt TriArc}. These degeneracies can be hard to break, and highlight the need to explore other wavelength bands. By retrieving at the next best band (that they do not share) each molecule can be distinguished and further information gathered. We therefore present the detection thresholds at multiple bands in \hyperref[fig:G395M]{Figure 5}, with the non-best-case bands serving as disambiguation thresholds.

In the case of the Hycean planet for example, the \ch{C2H2} detection threshold is 11 ppm, but the disambiguation threshold is 180 ppm. Therefore, for between 11--180 ppm only a detection at 3 $\mathrm{\mu m}$ would be observed (at a $3\sigma$ level), so either \ch{HCN} or \ch{C2H2} could be responsible for the detection. With \textgreater 180 ppm of \ch{C2H2}, there would also be an observation at 3.7 $\mathrm{\mu m}$, and the presence of \ch{C2H2} could be confirmed. The band at 3.5 $\mathrm{\mu m}$ could be explored by integrating the posterior PDF to constrain the maximum abundance of \ch{HCN} present. \sed{The full retrievals however, finds that while HCN and \ch{C2H2} are slightly degenerate at their detection thresholds, increasing the abundance by 5x is sufficient to distinguish between HCN and \ch{C2H2}. Therefore we find that the TriArc method of finding disambiguation thresholds massively overestimates the disambiguation threshold, as it does not use information from multiple spectral features.}

This hydrogen cyanide-acetylene degeneracy is the most severe and difficult to break due to the spectroscopic similarity of these molecules, as predicted by \citet{sousa2019molecular}. There is actually a four-way degeneracy between HCN, \ch{C2H2}\, \ch{NH3}, and \ch{HC3N} at 3 $\mathrm{\mu m}$, but both \ch{NH3} and \ch{HC3N} have other strong spectral features that can be observed, so disambiguation is much easier. The other non-trivial degeneracy is between \ch{CH4} and \ch{CH2O}, which possess very similar spectra at NIR wavelengths. These can be distinguished with the detection or non-detection of the feature of \ch{CH2O} at 6 $\mathrm{\mu m}$ with MIRI. \sed{While the difference between the detection and disambiguation threshold likely varies between pairs of degenerate molecules, no pair that we consider are more degenerate than the highly spectrally similar HCN and \ch{C2H2}.}

It is also worth noting that we are also assuming that the abundances of these species are independent, but our background knowledge of the photochemical stability and geophysical plausibility of the molecules could also inform our priors and hence impact our analysis of degeneracies.

The challenge of distinguishing molecular spectral features is a known difficulty in the field of atmospheric characterisation, exemplified by the contested detection of phosphine in the atmosphere of Venus \citep{greaves2021phosphine,villanueva2021no}. Higher spectral resolution can also allow observers to break degeneracies \citep{tremblay2020detectability}. \sed{For all of our calculations we bin to a decreased resolution of R=100, while the G395M instrument is capable of achieving resolutions of R$\approx$1000. The G395H instrument can achieve an even higher resolution of R$\approx$3400, at the cost of a gap in the spectrum useful for detecting \ch{H2S} and distinguishing \ch{C2H2} and HCN.}

\section{Discussion}
\label{sec:Discussion}

We have shown that all the prebiosignature molecules we have considered are detectable in hydrogen-rich exoplanets using a modest amount of observation time with \textit{JWST}. Cyanoacetylene (\ch{HC3N}) and formaldehyde (\ch{CH2O}) are the most readily detected primary prebiosignatures. Secondary prebiosignatures, \ch{CH4} and \ch{C2H2}, are also particularly well-suited to detection, and primary prebiosignatures \ch{SO2}, HCN, CO, and \ch{NH3} are detected in trace abundances in most cases. All of these are detected at wavelengths explored with the NIRSpec G395M instrument. Adding in the NIRISS SOSS instrument allows for the detection of \ch{H2S} (in moderate to large abundances) and can aid in the disambiguation of \ch{NH3}. Including observation with the MIRI LRS instrument allows the detection of NO (only in large abundances) and is also necessary to distinguish between \ch{CH2O} and \ch{CH4}. 

We present a discussion of the chemical and physical contexts which affect prebiosignatures for each planet in \hyperref[subsec:planets]{Section 4.1}. We discuss the impact of observational parameters and the application to observing strategies in \hyperref[subsec:observation]{Section 4.2}. We also touch on how these can be related to prebiotic chemistry experiments in \hyperref[subsec:prebiotic]{Section 4.3}.

\subsection{Impact of Planetary Properties on Detection Thresholds}
\label{subsec:planets}

The primary impact of planetary properties on the detection thresholds is through the scale height. This trend is most readily observed in the detection thresholds for \ch{CH4}, as its strong broad band opacity means that it will tend to dominate the transmission spectrum of a planet even in small abundances and avoid being obscured by absorption lines from another molecule. However, for most molecules the detection threshold is significantly lower in more transparent atmospheres with less broadly and strongly absorbing molecules like \ch{CO} (10 Myr Post-Impact planet) and \ch{CO2} (Super-Earth) than in the other \ch{CH4}-dominated atmospheres. The combination of \ch{CH4} and HCN in particular makes the detection thresholds for the ultrareduced volcanic planet particularly high. 

Clouds and hazes can also significantly impact detection thresholds by reducing the strength of absorption features. When decoupled from the assumed planetary radius, a grey cloud deck has an identical impact on detection thresholds as surface pressure. The sensitivity analysis of the impact of the pressure of the cloud deck demonstrates a steep dependence of detection threshold on cloud top pressure (or equivalently surface pressure) if they are found above the photosphere at approximately 10 mbar (see \hyperref[sec:sensitivity]{Appendix A}). In addition to water clouds, prebiosignatures may impact their own observability, through the generation of both photochemical hazes from HCN \citep[such as on Titan, e.g.][]{lara1999titan} and sulfur aerosols \citep{hu2013photochemistry}.

\subsubsection{Sub-Neptunes and Ocean Planets}

As sub-Neptunes are a particularly abundantly discovered type of exoplanet, it would be statistically very advantageous to any prebiosignature (or biosignature) survey if they do constitute a suitable environment for life  \citep{madhusudhan2021habitability}. The detection of water vapour in a low mean molecular weight atmosphere around sub-Neptune K2-18b \citep{benneke2019water,tsiaras2019water} is consistent with a range of internal compositions, including a water-rich composition \citep{madhusudhan2020interior}. That the planet does indeed have an ocean surface is not clear, \citet{scheucher2020consistently} exclude the existence of an oceanic surface due to the low mean molecular weight of the atmosphere and assumed evaporation from a water surface. For a planet such as K2-18b, it may require a thinner hydrogen envelope, strong cloud cover, or lower instellation to prevent sufficient water evaporation at the surface. JWST-detectable observational discriminants for the existence of a ocean surface beneath a thin envelope for sub-Netpune planets are explored by \citet{tsai2021inferring}. Notably, they find lack of \ch{NH3}, detectable with three transits of NIRSpec G395H, would be evidence for lack of an ocean surface. This makes the detection threshold for \ch{NH3} in a Hycean atmosphere that we have found particularly significant in this case, as \ch{NH3} is also both a prebiosignature, and a potential biosignature in hydrogen-dominated atmospheres \citep{seager2013biosignature}. Interestingly, we find the detection threshold of \ch{NH3} in a Hycean-type atmosphere to be lower at the wavelengths explored by the \sed{NIRISS SOSS} and G140M instruments, which is helpful to distinguish it from molecules with their dominant feature at 3 $\mu$m (like HCN and \ch{C2H2}). 

Another consideration for prebiosignature detection on a Hycean planet and other oceanic planets, is the impact of planet-wide oceans on the atmospheric chemistry. Soluble molecules, such as \ch{NH3}, \ch{CH2O}, and methanol (\ch{CH3OH}) would have difficulty accumulating in the atmosphere \citep[e.g.,][]{zhan2022organic,huang2022methanol}. Very large production rates would be required for prebiotically-relevant concentrations to develop in the oceans and, depending on the ocean-atmosphere equilibrium, no observable abundance of gas will accumulate in the atmosphere. \cite{pinto1980photochemical} demonstrate how photochemically-derived formaldehyde could accumulate in Earth's primitive oceans. The solubility of these species is considered in \hyperref[subsec:prebiotic]{Section 4.3} to determine what concentrations may build-up. \citet{tsai2021inferring} use \ch{CH3OH} as an observational discriminant between a solid and liquid ocean surface in K2-18b, observable with 20 \textit{JWST} transits using MIRI LRS. Many conventional environments proposed for the origin of life, such as tidal pools \citep{deamer1997first}, impact craters \citep{chatterjee2016symbiotic}, carbonate-rich lakes \citep{toner2020carbonate}, surface hydrothermal vents \citep{rimmer2019origin}, and hot springs \citep{damer2020hot}, couldn't occur on a planet completely lacking a solid surface.

However, providing sub-Neptunes with liquid water surfaces exist \sed{and are not uniformly impacted by high continuous cloud cover}, they greatly add to the sample of planets available to prebiosignature (and biosignature) analysis. They are well suited to detection due to their high scale height, although broad \ch{H2O} and \ch{CH4} features do increase the detection thresholds, particularly of \ch{HCN} and \ch{NO}. The presence of an ocean surface can be explored simultaneously to prebiosignatures by the detection or non-detection of \ch{NH3}. 

\subsubsection{Super-Earths and Volcanic Planets}

Super-Eaths are another abundantly detected class of exoplanet. A favoured explanation for the bi-modal distribution (radius gap) in planetary radii in super-Earth/sub-Neptune mass planets is the presence, or lack of, thick primary envelopes, driven by photoevaporation \citep[e.g.][]{lammer2003atmospheric} or core-powered mass loss \citep[e.g.][]{ginzburg2016super}. Planets that have retained their thick hydrogen-envelope are the sub-Neptunes discussed above. Considering planets without a thick hydrogen-envelope, we can still expect some super-Earths to possess secondary atmospheres from outgassing during accretion, or later due to tectonic processes \citep{elkins2008ranges,liggins2020can}. The higher gravity could allow a planet to retain hydrogen in its atmosphere \citep{miller2008atmospheric}, and significant hydrogen in-gassing during accretion could buffer mass loss with subsequent out-gassing \citep{chachan2018role}. Sulfur-species released by volcanism could cause significant aerosol cover that impacts detectability of other features, but in itself could be an observational discriminant for active volcanism and the presence of prebiosignatures \ch{H2S} and \ch{SO2} \citep{hu2013photochemistry,jordan2021photochemistry}.

Our modelling of a hydrogen-rich Super-Earth is performed assuming a stable hydrogen- and nitrogen-rich secondary atmosphere. Volcanism can cause carbon species like \ch{CO2} to accumulate to significant quantities, we consider only 100 ppm of \ch{CO2}, but both CO and \ch{CH4} could accumulate depending on the nature of the outgassing. If thin hydrogen-rich atmospheres are indeed a common feature of super-Earths, they are ideal candidates for the detection of prebiosignatures, particularly if the primary carbon species is \ch{CO2} or CO, which do not significantly interfere with the detection of most prebiosignature molecules. Even if enriched with higher mean molecular weight components due to hydrogen escape, the sensitivity analysis (\hyperref[sec:sensitivity]{Appendix A}) shows that prebiosignatures remain detectable for hydrogen abundances of around 70\%.

The ultrareduced volcanic planet we consider  \citep[e.g.,][]{swain2021detection}, relies on large quantities of hydrogen-rich outgassing to maintain the hydrogen-rich atmosphere, as the hydrogen would otherwise rapidly escape due to the planet's Earth-like gravity. Whether these observations prove to support the presence of a low mean molecular weight atmosphere with HCN is matter of debate \citep{mugnai2021ares,libby2022featureless}, so the physical relevance of the detection thresholds remains to be seen. However, if confirmed with \textit{JWST} observations, it highlights volcanism on some planets as a potential rich source of prebiotically relevant chemicals, detectable on a planetary scale. Both the observations of \citet{swain2021detection} and modelling of \citet{rimmer2021detectable} suggest the presence of aerosols or hazes composed of carbon-nitrogen species like on Titan \citep{clarke1997titan} which would further raise the already high detection thresholds. The surface temperature on GJ 1132b is also likely to be above that suitable for prebiotic chemistry. It is therefore highly uncertain how significant this hypothetical class of planet is to an overall study of the origin of life on exoplanets.

\subsubsection{Post-impact Planets}

The inherent transience of an impact-derived atmosphere makes it unclear how frequently we could expect to detect post-impact planets. The exact nature of the post-impact atmosphere is also sensitive to both the initial size and composition of the impacted planet, and the size and composition of the impactor, as well as the time since the impact \citep{zahnle2020creation}, but can be generalised as either recent (\ch{CH4}-dominated) or late (CO-dominated). Subsequent atmospheric evolution also depends on interactions of the impact-generated atmosphere with the now molten surface, which can lead to less reducing atmospheres than the ones we used in our prebiosignature analysis \citep{itcovitz2022reduced}.

The recent post-impact condition (represented in our models by the 100 kyr post-impact planet, but the exact length of the time the condition persists is sensitive to many parameters, especially stratospheric \ch{H2O} abundance) is extremely well suited to prebiotic chemistry, with prebiotic chemicals like \ch{HCN} and \ch{CH4} present in the atmosphere. However, it suffers from the issue that the surface temperature will likely exceed those allowed for most prebiotic syntheses. Furthermore detectability thresholds are likely to be reduced by photochemical hazes and aerosols, and this post-impact state doesn't last long so may not be well-represented in the exoplanet population.

The later post-impact state (represented in our models by the 10 Myr post-impact planet), where the atmosphere is predominantly \ch{H2} and CO, is exceptionally good for the detection of trace species. With a carbon to oxygen (C:O) ratio of approximately unity, prebiotic species like HCN can be formed from lightning, impacts and stellar activity \citep{rimmer2019hydrogen}. This state also persists for a considerable length of time, with steadily decreasing abundances of hydrogen due to escape processes, favouring the discovery of an exoplanet in such a state. The surface temperature is likely to be more temperate than in the earlier atmospheric evolution, but may still be hostile to prebiotic chemistry due to the greenhouse effect of \ch{H2} \citep{pierrehumbert2011hydrogen}. Overall, if we discount their possible rarity, post-impact planets are ideal targets for detecting prebiosignatures.

\subsubsection{Early Earth}

High mean molecular weight atmospheres on exoplanets around red dwarfs are not beyond the realm of study of \textit{JWST}, but require significant observation time. There are also very few known planetary systems amenable to such a study \citep{morley2017observing}. The most promising candidate planet, TRAPPIST-1e, has accessible detection thresholds for \ch{CH4}, \ch{NH3}, HCN, \ch{HC3N}, and \ch{C2H2} with 5-10 transits, but other prebiosignatures require a perhaps prohibitively long observation program (40-100 transits). Should initial observations of TRAPPIST-1e or another TRAPPIST-1 planet prove fruitful, we may find ourselves devoting significant observation time to the TRAPPIST-1 system. Beyond this system we will need to focus on finding prebiosignatures in hydrogen-rich atmospheres (with \textit{JWST} at least), rather than early-Earth or other solar system  atmosphere analogues. The study of early Earth atmospheres in exoplanets is a task better suited to future observatories.

\subsection{Impact of Observing Strategy on Prebiosignature Detection}
\label{subsec:observation}

The detection thresholds we have calculated are sensitive to the duration of observing time, as well as the radius and magnitude of the observed star. The detection threshold decreases with the number of transits at a roughly constant rate in log-log space. The \sed{slope of the dependence on detection threshold on the number of transits} depends on the species in question (\hyperref[fig:transitshycean]{Figure 9}), but largely does not depend on the star \sed{when considering M dwarfs} (\hyperref[fig:transitsstars]{Figure 10}). A systematic noise floor, such as that found by \citet{rustamkulov2022analysis}, will cause the detection threshold to eventually level off, which will happen with fewer transits for brighter stars. Prebiosignatures in hydrogen-rich atmospheres are feasibly detected in five transits in M dwarfs of 10\textsuperscript{th} J magnitude or less. Brighter late K dwarfs are feasible targets, as are smaller and dimmer M dwarfs, notably including TRAPPIST-1 and Kepler-1649 \citep{vanderburg2020habitable}. The precise number of transits that is ideal for any particular planetary system will depend on the star, the planetary properties, and the desired detection threshold. For example, high mean molecular weight atmospheres are moderately well characterized in the case of very small stars (e.g. TRAPPIST-1) after ten transits, but detection of all prebiosignatures in these atmospheres requires one hundred transits (\hyperref[fig:thresholdstrappist]{Figure 11}).

By far the most important \textit{JWST} instrument for prebiosignature detection is NIRSpec G395M (or G395H), as the most important spectral features for \ch{CH4}, \ch{C2H2}, \ch{HC3N}, HCN, CO, \ch{SO2}, and \ch{NH3} all exist in the 2.9--5 $\mathrm{\mu m}$ range. A more conservative observational regime could use NIRSpec G395M exclusively (potentially combined with a single transit of NIRISS SOSS to enhance atmospheric retrievals). This would sacrifice the ability to detect NO (which requires MIRI LRS) and \ch{H2S} (which requries NIRSpec Prism, NIRISS SOSS, or NIRSpec G140M), and to disambiguate molecules like \ch{CH4} from \ch{CH2O}, and potentially \ch{HCN} from \ch{NH3}. For the 1--2.9 $\mathrm{\mu m}$ range, NIRISS SOSS, or NIRSpec Prism should be used depending on the brightness of the target. The wider spectral baseline of these instruments makes them more attractive than using the NIRSpec G140M instrument when performing retrievals, even though we do not find any detection or disambiguation thresholds in the 2--2.9 $\mathrm{\mu m}$ range.

\subsection{Prebiotic Consequences of the Detection Thresholds}
\label{subsec:prebiotic}

The prebiosignatures that we focus on here are relevant for prebiotic chemistry that takes place in water. The ``success'' of the prebiotic chemistry depends critically on the concentration of these species in liquid water, and there is a close but complex relationship between atmospheric and surface water concentrations. This relationship depends on the local geochemistry of these waters. Drawing the connection between global atmospheric partial pressures of a molecule and expected global and local concentrations of that same molecule is outside the scope of this paper.

Aqueous concentrations are most relevant for {\bf primary prebiosignatures}: species that participate directly in the chemical synthesis of prebiotically relevant compounds. The atmospheric partial pressures of {\bf secondary biosignatures}: species that indicate events and/or environmental factors that may be conducive for certain prebiotic chemical scenarios (e.g., lightning, giant impacts, volcanism), are related to the ubiquity and intensity of these processes, and these relations have been worked out for several secondary prebiosignatures \citep[e.g.,][]{kaltenegger2009detecting,rimmer2019hydrogen,rimmer2019identifiable,rimmer2021detectable,rimmer2021life}.

\section{Conclusion}
\label{sec:Conclusion}

We have described the creation of a pipeline to compute the minimum abundance of a molecule required in an exoplanet atmosphere to be detected by \textit{JWST}. Using this pipeline we have have computed the \textit{JWST} detection thresholds for ten prebiosignature molecules in a selection of model exoplanets. We have varied the observational regime by choosing different stars and number of transits to quantify the effect this has on the detection thresholds. We discussed how these results relate to the relevance of prebiosignature detection with each model exoplanet we used and how they could inform observing strategies when attempting to detection prebiosignatures.

Our key finding is that the study of prebiosignatures and the origin of life in an exoplanetary context is well within the capabilities of \textit{JWST}, in the case of a low mean molecular weight atmosphere or an optimal target system (like TRAPPIST-1). All ten of our prebiosignatures are detectable in all the hydrogen-rich atmospheres we consider, using a modest number of transits ($<$5) and applicable to a reasonable number of target stars. Notably, quantities of HCN, \ch{H2S}, or \ch{HC3N}, species at the heart of the cyanosulfidic and other prebiotic scenarios, are well detected as atmospheric species. We can therefore directly constrain these scenarios with observations. Secondary prebiosignatures \ch{C2H2}, \ch{CH4}, and CO are frequently detectable in very low abundances of the order of a few ppm, which may help constrain both prebiotic chemistry, atmospheric redox, and other relevant processes including impacts and volcanism. We also have well-constrained detection thresholds for gaseous \ch{SO2}, NO, and \ch{NH3}, which will be strongly related to the aqueous chemistry of surface waters. For the most part, we find that high mean molecular weight atmospheres (such as that of early Earth) are not suitable to the detection of prebiosignatures, but we do find that in the case of TRAPPIST-1e, many prebiosignatures are detectable with ten transits. TRAPPIST-1, and similar ultracool stars like SPECULOOS-2 \citep{delrez2022two}, are therefore our most immediate avenue with which we might explore Earth-like atmospheres. The tools we have developed in this paper give the opportunity to repeat this kind of detection threshold analysis for any situation where detection of a trace amount of an atmospheric species would be interesting. Beyond prebiosignatures, this has obvious applications to biosignatures and other atmospheric processes.

\textit{JWST} has proven a powerful tool in the remote sensing of exoplanet atmospheres, having already made novel discoveries of \ch{CO2} \sed{and \ch{SO2}} in the atmosphere of WASP-39b \citep{jwst2022identification,ahrer2023early,alderson2023early,feinstein2023early,rustamkulov2023early}. \sed{Notably the detection of \ch{SO2} is evidence for photochemistry \citep{tsai2022direct}, and is a prebiosignature in appropriate planetary contexts.}

Furthermore, many of the planets we have discussed as suitable for prebiosignature analysis, including the TRAPPIST-1 planets, GJ 1132b, LTT 1445 Ab, and K2-18b are all targets of the \textit{JWST} GO Cycle 1 or GTO programs. We therefore expect a wealth of data about the atmospheres of these planets, which may even include the detection of prebiosignature molecules, and should certainly help understand the atmospheric contexts present in terrestrial planets. Follow up observations to further characterise the atmosphere may yield highly useful data of great prebiotic relevance, able to confirm or rule out the presence of prebiosignatures. Even if \textit{JWST} observations of terrestrial planets yield flat spectra, as seen in recent shorter-wavelength observations of GJ 1132b \citep{libby2022featureless}, LTT 1445 Ab \citep{diamond2022ground}, and L98-59b \citep{damiano2022transmission}, we can still expect to study relevant atmospheric processes to the origin of life with \textit{JWST} by observation of temperate and warm planets with significant hydrogen envelopes. \sed{Looking beyond \textit{JWST}, the wavelength range that we consider also overlaps with the wavelengths considered by \textit{LIFE} (4-18 $\mathrm{\mu}$m), so detection of prebiosignatures in directly imaged Earth-like planets may be possible in the future.}\\

We thank the anonymous referee for their insightful feedback which have greatly improved the paper. The authors would like to thank J. Itcovitz for his feedback and data regarding post-impact atmospheres, and P. Wheatley for his comments and advice. ABC acknowledges support from the Institute of Astronomy and the UK Science and Technology Facilities Council (STFC). PBR thanks the Simons Foundation for support under SCOL awards 59963. SR acknowledges support from the Glasstone Foundation and of the Natural Sciences and Engineering Research Council of Canada (NSERC) Discovery Grant, [2022-04588].

\software{ {\tt petitRADTRANS} \citep{molliere2019petitradtrans},
{\tt PandExo} \citep{batalha2017pandexo}, 
{\tt TriArc}\footnote{https://github.com/ExoArcturus/TriArc},
{\tt exo-k} \citep{leconte2021spectral}}

\appendix
\restartappendixnumbering

\section{Sensitivity Analysis}
\label{sec:sensitivity}

\begin{figure}[p]
    \centering
    \includegraphics[width=0.45\textwidth]{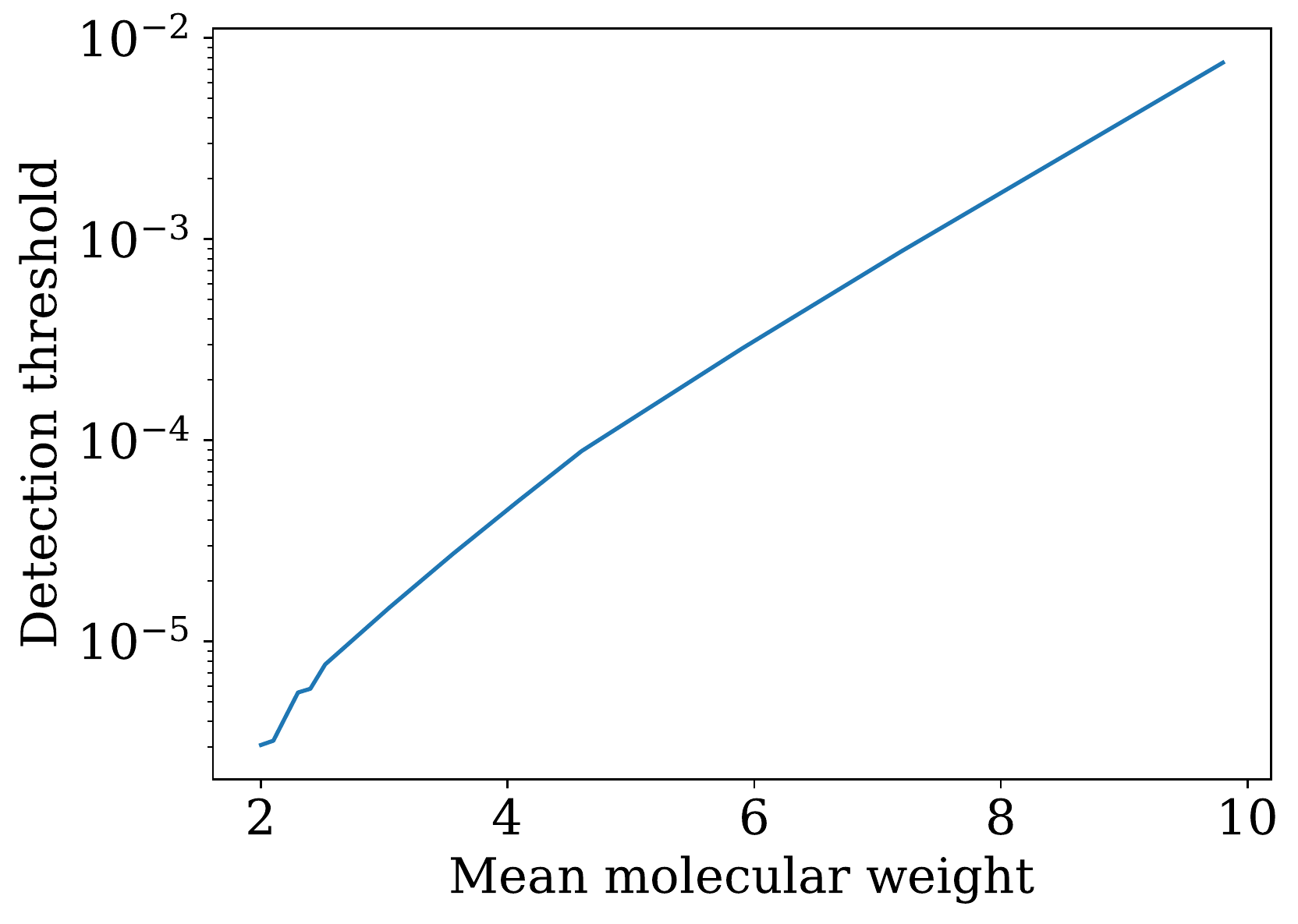}
    \includegraphics[width=0.45\textwidth]{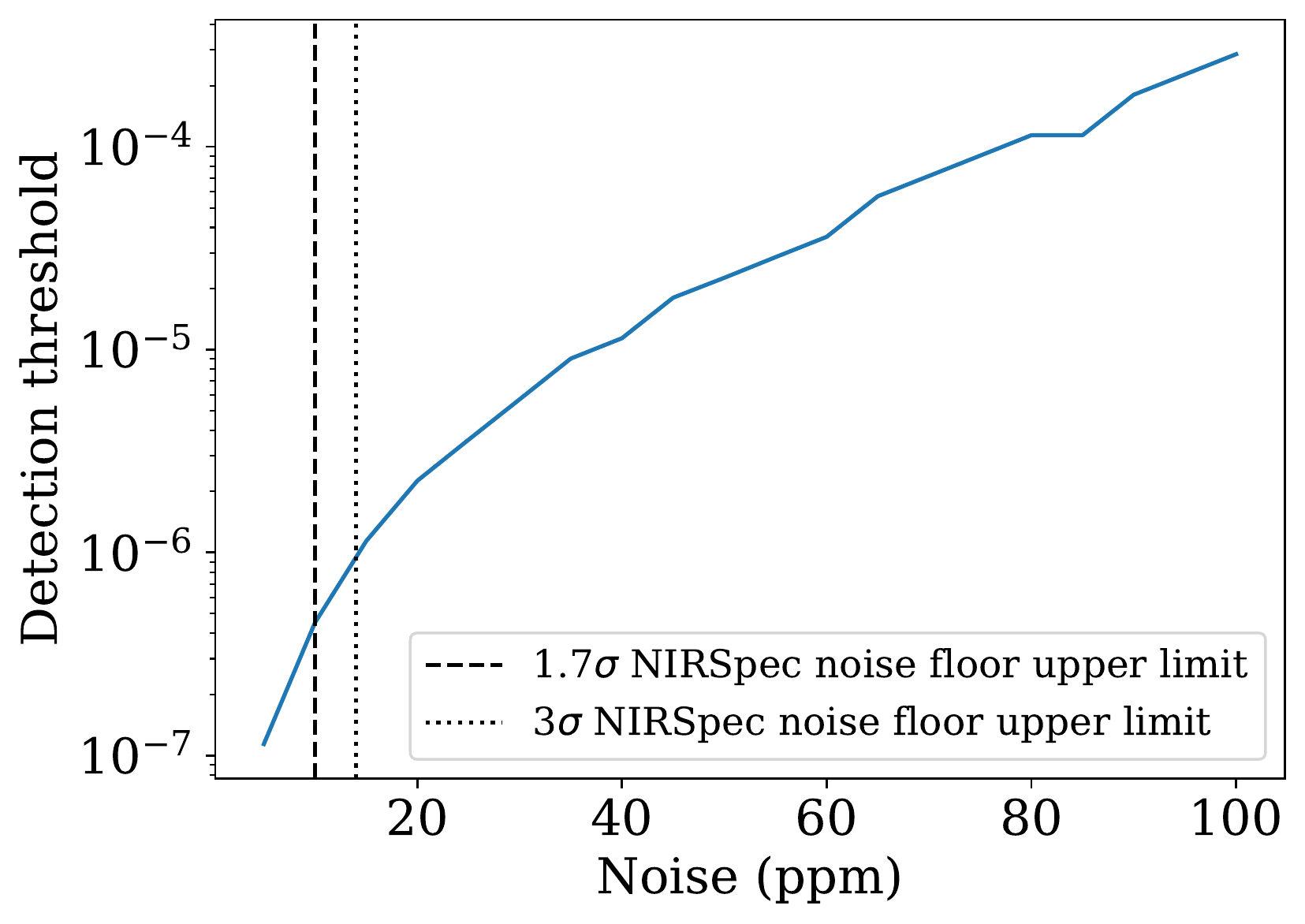}
    \includegraphics[width=0.45\textwidth]{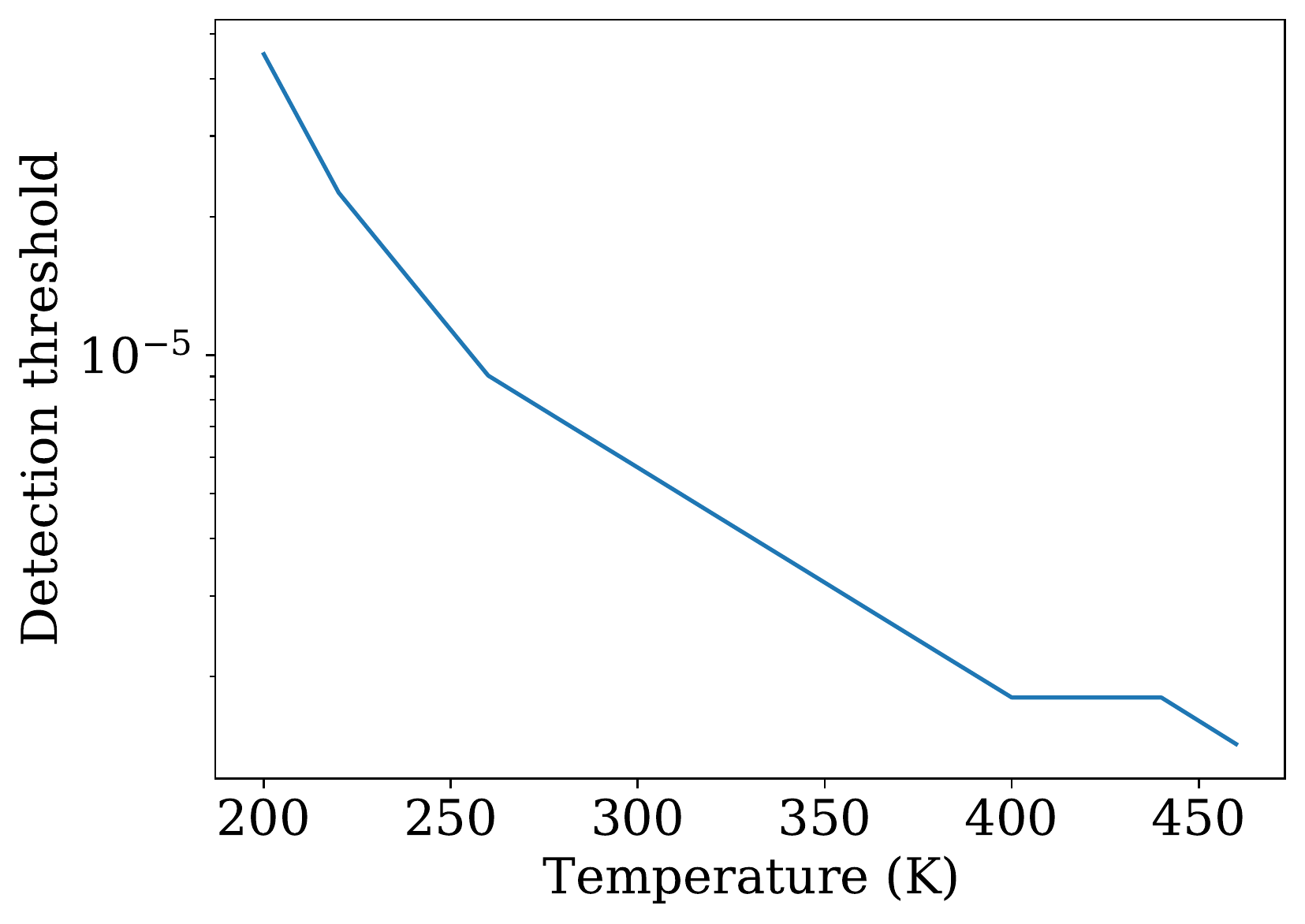}
    \includegraphics[width=0.45\textwidth]{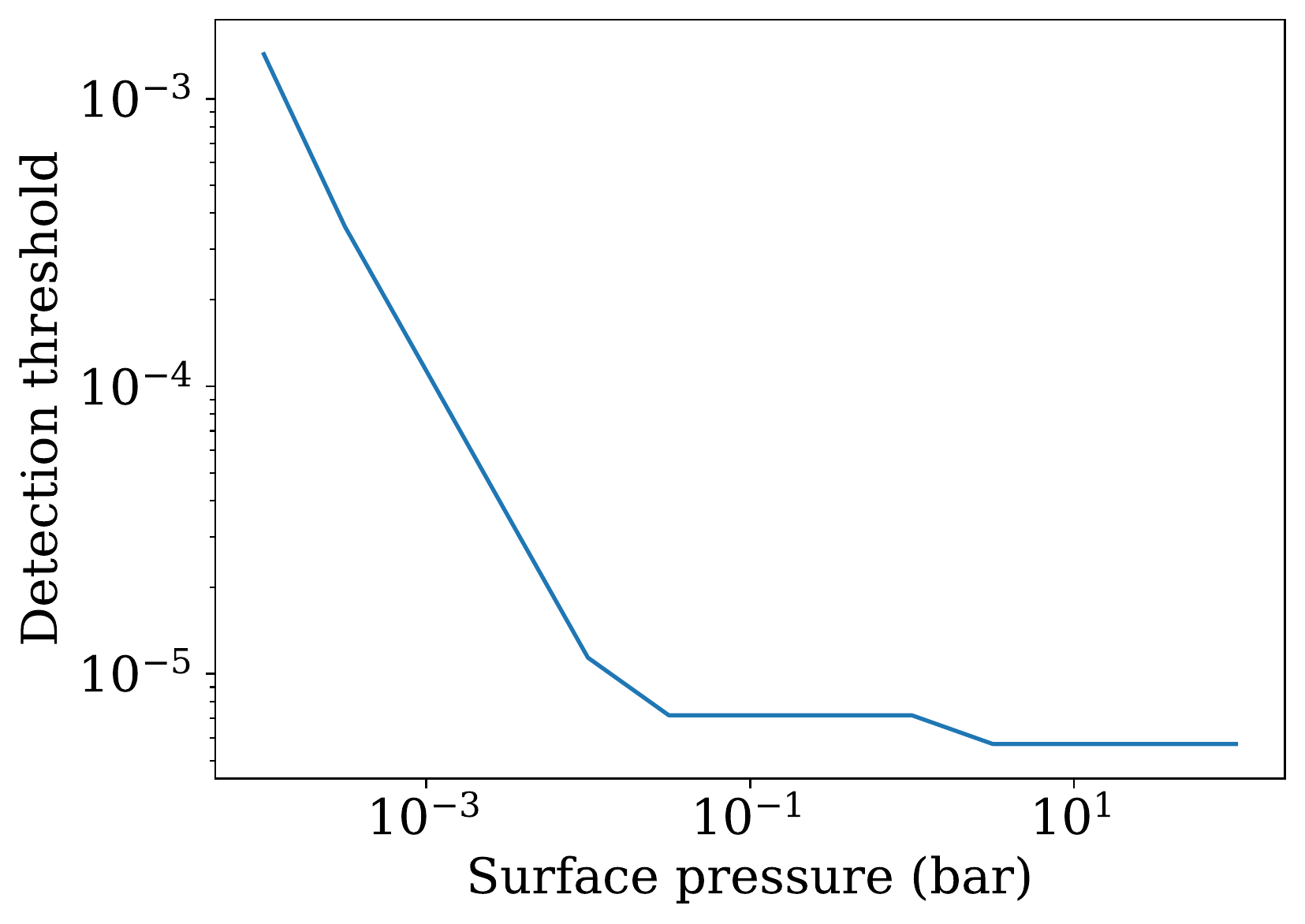}
    \includegraphics[width=0.45\textwidth]{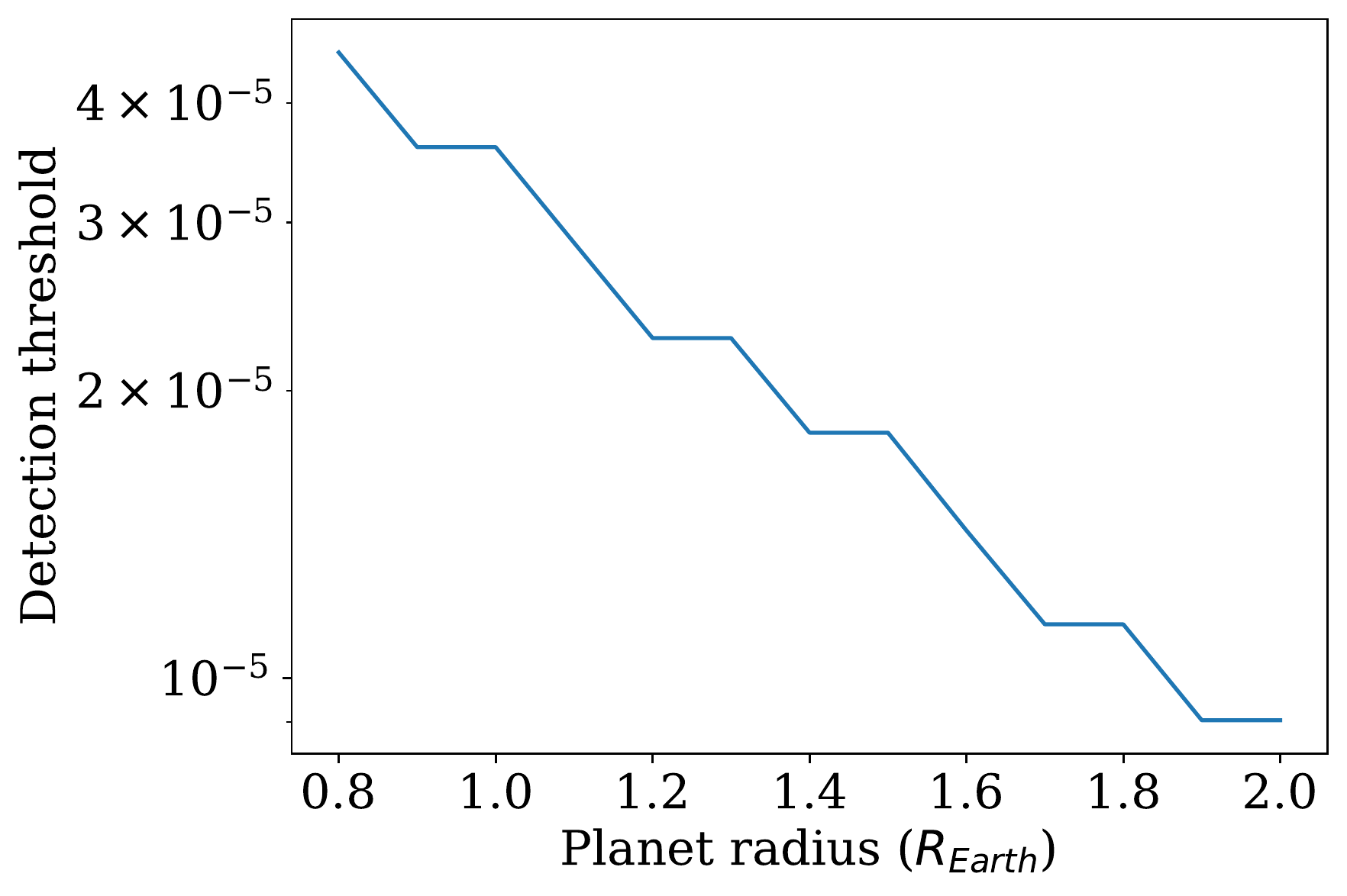}
    \includegraphics[width=0.45\textwidth]{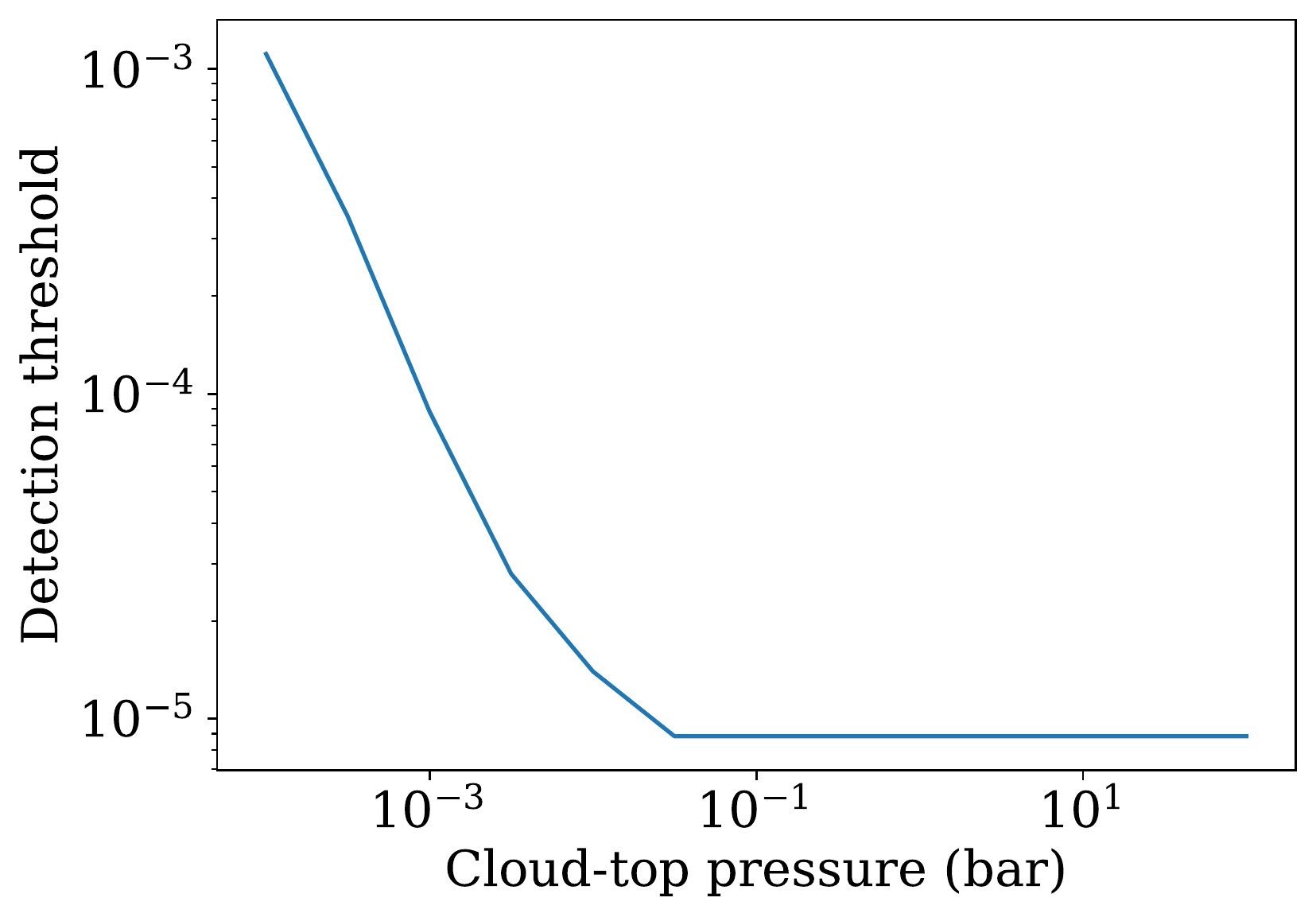}
    \caption{Detection threshold of \ch{C2H2} as a function of the atmosphere's mean molecular weight (top left), transmission spectrum noise (top right), temperature of the isothermal atmosphere (middle left), surface atmospheric pressure (middle right), the planet radius (bottom left), and the cloud top pressure (bottom right). In each case only a single parameter is varied, with other parameters given by a model Hycean planet with noise from observing GJ 1132b for three transits. The noise floors for NIRSpec is given by \citet{rustamkulov2022analysis}.}
    \label{fig:sensitivity}
\end{figure}

In order to explore the sensitivity of the detection thresholds to single key parameters, we use the detection of \ch{C2H2} at 3 $\mathrm{\mu m}$ as a benchmark. In each case we vary a single parameter, holding all others as the constant values described for the model Hycean planet with three transits around GJ 1132b. The parameters that varied are mean molecular weight (by changing the fraction of non-absorbing species \ch{H2}, He, \ch{N2}), isothermal atmosphere temperature, instrument spectral precision, surface pressure, and planet radius. Mean molecular weight and temperature (along with surface gravity) determine the scale height, which along with the planet radius, determines the signal strength of spectral features. The high sensitivity of the detection thresholds to these parameters can be seen in \hyperref[fig:sensitivity]{Figure A1}, varying by a factor of 10 over the range of reasonable temperatures, and a factor of 5 over the range of reasonable planetary radii. The importance of mean molecular weight (and hence hydrogen abundance), which can reasonably range from two to 24, is highlighted.

The sensitivity analysis also demonstrates the importance of surface pressure, or cloud-top pressure in the case of a grey opaque cloud deck. Above the photosphere, found in this case at about 10 mbar, the detection threshold is constant and doesn't depend on surface/cloud-top pressure. In the case of a thin atmosphere, with a surface/cloud-top pressure of less than 10 mbar, the detection threshold decreases steeply with decreasing surface pressure. Therefore both high cloud and thin atmospheres can both significantly impact detection thresholds by suppressing spectral features \citep[e.g.][]{kreidberg2014clouds}. The strong dependence of detection threshold on instrument noise is responsible for the improvements that come from observing increasing numbers of transits, with the \textit{JWST} noise floor upper limits found by \citet{rustamkulov2022analysis} added.

\bibliography{biblio}

\end{document}